\documentclass[preprint,12pt,authoryear]{elsarticle}
\usepackage{microtype}
\usepackage{amssymb}
\usepackage[utf8]{inputenc}   
\graphicspath{ {./figures/} }
\usepackage{hyperref}
\usepackage{floatrow}
\usepackage{verbatim} 
\usepackage{amsfonts}
\usepackage{booktabs}
\usepackage{threeparttable}
\usepackage{tabularx}
\usepackage{makecell}
\usepackage{hyperref}
\usepackage{lineno}
\usepackage{amsmath}
\usepackage{algorithm,float}
\usepackage{algorithmicx}
\usepackage{algpseudocode}
\usepackage{arydshln}
\usepackage{subcaption}
\usepackage{multirow}
\usepackage{enumitem} 
\usepackage{threeparttable}    
\makeatletter
\newenvironment{breakablealgorithm}
  {
   \begin{center}
     \refstepcounter{algorithm}
     \hrule height.8pt depth0pt \kern2pt
     \renewcommand{\caption}[2][\relax]{
       {\raggedright\textbf{\ALG@name~\thealgorithm} ##2\par}%
       \ifx\relax##1\relax 
         \addcontentsline{loa}{algorithm}{\protect\numberline{\thealgorithm}##2}%
       \else 
         \addcontentsline{loa}{algorithm}{\protect\numberline{\thealgorithm}##1}%
       \fi
       \kern2pt\hrule\kern2pt
     }
  }{
     \kern2pt\hrule\relax
   \end{center}
  }
\makeatother

\journal{Expert Systems with Applications}

\begin{document}

\floatsetup[table]{capposition=top}

\begin{frontmatter}

\title{iHQGAN:A Lightweight Invertible Hybrid Quantum-Classical Generative Adversarial Networks for Unsupervised Image-to-Image Translation}

\author[lable1,lable2]{Xue Yang}
\author[lable1,lable2]{Rigui Zhou$^\dagger$}
\author[lable1,lable2]{ShiZheng Jia}
\author[lable1,lable2]{YaoChong Li}
\author[lable1,lable2]{Jicheng Yan}
\author[lable1,lable2]{ZhengYu Long}
\author[lable1,lable2]{Wenyu Guo}
\author[lable1,lable2]{Fuhui Xiong}
\author[lable1,lable2]{Wenshan Xu}

\affiliation[lable1]{organization={School of Information Engineering, Shanghai Maritime University},
            city={Shanghai},
            postcode={201306}, 
            country={China}}
\affiliation[lable2]{organization={Research Center of Intelligent Information Processing and Quantum Intelligent
Computing},
            city={Shanghai},
            postcode={201306},      
            country={China}}      

\begin{abstract}
Leveraging quantum computing’s intrinsic properties to enhance machine learning has shown promise, with quantum generative adversarial networks (QGANs) demonstrating benefits in data generation. 
However, the application of QGANs to complex unsupervised image-to-image (I2I) translation remains unexplored. Moreover,classical neural networks often suffer from large parameter spaces, posing challenges for GAN-based I2I methods.
Inspired by the fact that unsupervised I2I translation is essentially an approximate reversible problem, we propose a lightweight invertible hybrid quantum-classical unsupervised I2I
translation model — iHQGAN, by harnessing the invertibility of quantum
computing.Specifically,iHQGAN employs two mutually approximately reversible quantum generators with shared parameters, effectively reducing the parameter scale. To ensure content consistency between generated and source images, each quantum generator is paired with an assisted classical neural network (ACNN), enforcing a unidirectional cycle consistency constraint between them. Simulation experiments were conducted on 19 sub-datasets across three tasks. 
Qualitative and quantitative assessments indicate that iHQGAN effectively performs unsupervised I2I translation with excellent generalization and can outperform classical 
methods that use low-complexity CNN-based generators.
Additionally, iHQGAN, as with classical reversible methods, reduces the parameter scale of classical irreversible methods via a reversible mechanism.
This study presents the first versatile quantum solution for unsupervised I2I translation, extending QGAN research to more complex image generation scenarios and offering a quantum approach to decrease the parameters of GAN-based unsupervised I2I translation methods.
\end{abstract}


\end{frontmatter}

\section{Introduction}
\label{introduction}
Quantum computing \cite{preskill2012quantumcomputingentanglementfrontier,harrow2017quantum} is a new computational paradigm that leverages the unique properties of quantum mechanics.
Quantum machine learning (QML) combines quantum computing with machine learning, aiming to enhance the performance of machine learning(ML) by leveraging quantum computing's advantages. 
Currently, we are in the Noisy Intermediate-Scale Quantum (NISQ) era 
 \cite{preskill2018quantum}, characterized by a limited number of controllable qubits that are prone to noise. A key direction of QML is the hybrid quantum-classical models 
 \cite{callison2022hybrid} based on variational quantum algorithms(VQA) \cite{cerezo2021variational}, which are well-suited for NISQ devices and have shown remarkable performance in domains like data generation \cite{lloyd2018quantum}. 
Recently, QGANs have gained attention in quantum image generation and achieved significant milestones\cite{huang2021experimental,tsang2023hybrid,zhou2023hybrid,chu2023iqgan,silver2023mosaiq}. However, they are still in the early stages of development, and their application to more intricate image generation scenarios remains an area for further exploration.

One particularly challenging task in the field of image generation is unsupervised I2I translation \cite{zhu2017unpaired}.
Unsupervised I2I translation is an approximately reversible task that learns bidirectional mappings between two domains without paired training data.
It enables image transfer between the source and target domains while ensuring content consistency, such as structural and spatial consistency. 
Mathematically, the reversibility of I2I translation can be characterized using inverse functions. 
Specifically, this involves identifying a function 
$f$ that maps images from the source domain to the target domain, along with its inverse function $f^{-1}$, which maps images from the target domain back to the source domain.
Unsupervised I2I translation underpins the performance of various applications in many areas, such as super-resolution \cite{yuan2018unsupervised,zhang2019multiple}, image style transfer \cite{tomei2019art2real,richardson2021encoding}, and image inpainting \cite{zhao2020uctgan,huang2017coarse,demir2018patch}.
Classical methods often enhance the efficiency of image feature learning by incorporating computationally intensive modules such as attention mechanisms \cite{emami2020spa,tang2021attentiongan}, resulting in a significant increase in model parameters. Moreover, for multi-target I2I translation \cite{bhattacharjee2020dunit,lee2018diverse}, 
utilizing multiple generators to transfer images into different target domains also enlarges the parameter scale.

Existing methods for reducing parameters in unsupervised I2I translation can be broadly categorized into irreversible-based and reversible-based approaches. 
The former focuses on reducing the parameters of I2I translation models through techniques such as knowledge distillation \cite{bhattacharjee2020dunit,lee2018diverse}, neural architecture search (NAS) \cite{yang2021netadaptv2,van2019evolutionary}, or improvements to heavyweight modules \cite{deng2023involutiongan}. However, despite efforts to minimize the number of parameters, these models fail to leverage the approximate reversibility of unsupervised I2I translation.
For reversible-based methods, most of the work incorporates invertible neural networks (INNs) \cite{dai2021iflowgan,van2019reversible} to construct reversible models. 
Due to the high computational overhead of INNs,
they are commonly utilized as components within the overall framework. The reversible components typically 
handle data that has already undergone dimensionality reduction in the initial phases.
Additionally, there is a study \cite{shen2020one} that proposes a self-inverse network for unpaired image-to-image translation. This approach involves augmenting the training data by swapping inputs and outputs during the training process, along with implementing separated cycle consistency loss for each mapping direction.
Reversible methods can allow parameter sharing and thus reduce the number of parameters.
Inspired by this, we consider constructing a reversible model by utilizing the invertibility of quantum computing, offering a novel quantum solution to reduce the parameter scale.

In light of the challenges mentioned above,
our research presents an invertible hybrid quantum-classical unsupervised I2I translation model—iHQGAN.
By combining the approximate reversibility of unsupervised I2I tasks with the inherent reversibility of quantum computing, iHQGAN provides a quantum solution for unsupervised I2I translation, addressing the challenge of large parameter scales in GAN-based methods.

In iHQGAN, we designed two mutually approximately reversible quantum generators with shared parameters, effectively reducing the parameter scale.
To maintain content consistency between the input and output images, we deployed an assisted classical neural network (ACNN) for each quantum generator, ensuring the implementation of a unidirectional cycle consistency constraint between
them.
Our research contributions can be summarized as follows:
\begin{enumerate}
\item 
iHQGAN is the first versatile quantum method for unsupervised I2I translation. It integrates the approximate reversibility of unsupervised I2I translation with the invertibility of quantum computing, extending the research of QGANs to more complex scenarios and reducing the parameter scale of GAN-based unsupervised I2I translation methods.
\item
iHQGAN leverages the invertibility of quantum computing to design two mutually reversible quantum circuits with shared parameters, thereby effectively reducing the parameter scale.
\item
To preserve content consistency between the generated and source images,iHQGAN deploys an ACNN for each quantum generator, introducing a unidirectional cycle consistency constraint
between them.

\item
Simulation experiments were conducted on 19 sub-datasets across three tasks. Both qualitative and quantitative assessments indicate that iHQGAN effectively executes unsupervised I2I translation with robust generalization, and can outperform classical methods that use low-complexity CNN-based generators.
Additionally, iHQGAN, similar to classical reversible methods, reduces the parameter scale of classical irreversible methods by introducing the reversible mechanism.
\end{enumerate}

\section{Preliminaries}
\label{Background}
\subsection{Fundamentals of Quantum Computing}
\label{Quantum Preliminary}
The basic unit for computing and storing information in quantum computing is the quantum bit (qubit). 
The superposition state of a single quantum bit can be represented as $\left | \psi  \right \rangle = 
\alpha \left | 0  \right \rangle +  \beta \left | 1  
 \right \rangle$,where
$\alpha$ and $\beta $ are complex numbers called probability amplitudes, satisfying  $|\alpha|^{2}+|\beta|^{2}=1$. When quantum measurements are performed on $|\psi|$, with probability  $|\alpha|^{2}$ ,the state collapses to
$\left|0\right\rangle $, and with probability  $|\beta|^{2}$, it collapses to $\left|1\right\rangle $.
The entangled state means that the state of one qubit is inherently linked to the state of the other, regardless of the distance between them.
A typical example of an entangled state is $(|00\rangle+|11\rangle) / \sqrt{2}$, where there is a special correlation between two qubits such that no matter how far apart they are, measurements on one immediately affect the state of the other.

Quantum gates can manipulate quantum states. The most commonly used single-qubit gates are $R x(\theta)$, $R y(\theta)$, and $R z(\theta)$. A two-qubit gate can create entanglement between two qubits. For example, the controlled $NOT$ ($CNOT$) gate. The $CZ$  gate acts on two qubits, the control qubit and the target qubit.
The $ROT$ gate is defined as  $\operatorname{Rot}(\alpha, \beta, \gamma)=R z(\gamma) R y(\beta) R z(\alpha)$.In this paper, the symbol $R$  is used to denote the $ROT$ gate.All quantum gates are unitary.

The unitary nature of quantum system evolution inherently guarantees the invertibility of quantum computations. According to the fundamental principles of quantum mechanics, the time evolution of a quantum system is described by the Schrödinger equation, which governs the evolution of the state vector$|\psi(t)\rangle$ from an initial time $t_0$ to a later time $t$:
$i h \frac{\partial}{\partial t}|\psi(t)\rangle=\widehat{H}|\psi(t_0)\rangle$.
As time evolves from $t_0$ to $t$, the quantum state transitions from $|\psi(t_0)\rangle$ to $|\psi(t)\rangle$. This evolution can be expressed using a unitary operator $U(t, t_0)$, which is given by:
$
|\psi(t)\rangle={U}\left(t, t_0\right)\left|\psi\left(t_0\right)\right\rangle$,
where $U(t, t_0) = e^{-i\hat{H}(t-t_0)/\hbar}$ is a unitary operator satisfying $U^\dagger U = UU^\dagger = I$. Due to the unitarity of this operation, the initial state can also be reconstructed from the final state as follows:
$|\psi(t_0)\rangle={U^{\dagger}}\left(t, t_0\right)\left|\psi\left(t\right)\right\rangle
$.

Variational quantum algorithms (VQA) solve specific problems by optimizing the parameters of parameterized quantum circuits (PQC) to minimize a target function. 
A PQC is composed of quantum registers, parameterized quantum gates, and measurement operations. 
In a PQC, the first step is to prepare the initial quantum state $|\alpha\rangle$ for input. Then, a series of unitary operations (quantum gates) is applied, and the final output quantum state is obtained through a measurement operator $\hat O$. A PQC  can be represented as :  
\begin{equation}
f(\theta)=\left\langle\alpha\left|U^{\dagger}(\theta) \hat{O} U(\theta)\right| \alpha\right\rangle
\end{equation}
Where $\theta=\left(\theta_1, \theta_2, \cdots\right)$ is the set of parameters of the quantum gate, $U$ represents the unitary transformation of the quantum gates, $\hat{O}$ is the measurement operator and $f(\theta)$ is the loss function defined according to the specific task.

\subsection{Quantum Generative Adversarial Networks(QGANs)}GANs were first introduced by Goodfellow et al. in 2014  \cite{goodfellow2020generative}. The key idea of GANs is to achieve a Nash equilibrium between the generator and the critic by having them compete with each other, compelling the generator to produce images that the critic cannot distinguish from real images. Primitive GANs often suffer from difficult convergence and mode collapse problems. To address these issues, 
researchers have proposed various variants. Notably, WGANs \cite{pmlr-v70-arjovsky17a} introduced the Wasserstein distance as a novel cost function. Building on this, WGANs-GP \cite{DBLP:journals/corr/GulrajaniAADC17} introduced the gradient penalty as a replacement for weight clipping, enforcing the implementation of the Lipschitz constraint to further enhance training stability compared with WGANs.GANs have achieved impressive results in unsupervised I2I translation \cite{yuan2018unsupervised,zhang2019multiple,tomei2019art2real,huang2017systematic,zhao2020uctgan,huang2020feature,demir2018patch}.

Inspired by GANs, Lloyd et al.\cite{lloyd2018quantum} introduced the concept of QGANs in 2018, integrating quantum computing mechanisms into the GAN framework. They demonstrated that QGANs exhibit an exponential advantage over classical methods when using data consisting of samples from measurements made in high-dimensional spaces.
Like classical GANs, QGANs also include generators and critics  trained against each other. Lloyd et al.\cite{lloyd2018quantum}
emphasized that the generator must be quantum, whereas the critic can be either classical or quantum.
Dallaire-Demers et al.\cite{dallaire2018quantum} successfully trained a quantum equivalent of the conditional GAN, named QuGAN, thereby validating the theory proposed by Lloyd et al. 
The objective function of QGANs often leverages classical distance metrics, such as Wasserstein distance \cite{tsang2023hybrid}, and binary cross-entropy 
\cite{huang2021experimental,zoufal2019quantum,situ2020quantum}.
When both the generator and critic are quantum in QGAN, objective functions can also use measures of distance between quantum states, including quantum state fidelity \cite{stein2021qugan}, quantum Wasserstein distance \cite{chakrabarti2019quantum}, and quantum Rényi distance \cite{kieferova2021quantum}.
Recently, QGANs have demonstrated successful applications in producing diverse data including  target quantum states \cite{benedetti2019adversarial},
electrocardiogram (ECG) data \cite{qu2024hq}
, chemical small molecules \cite{li2021quantum}, and images generation 
\cite{li2021quantum,huang2021experimental,zhou2023hybrid}.
Currently, QGANs are still in the early exploratory stage in the field of image generation. Due to limitations in quantum resources, researchers often employ dimensionality reduction techniques, such as principal component analysis (PCA) \cite{silver2023mosaiq} or autoencoders \cite{chang2024latent}, during the preprocessing phase of image data. The low-dimensional data generated by the quantum generator is used in the post-processing phase to reconstruct high-dimensional data.
Recently, Huang et al. \cite{huang2021experimental} introduced a quantum patch strategy that segments images into smaller patches, utilizing multiple quantum generators to independently generate the corresponding patches, which are then combined to form a complete image. This method effectively leverages the limited number of qubits, enabling the generation of 8x8 images without the need for dimensionality reduction. Building on this, Tsang et al. \cite{tsang2023hybrid} proposed PQWGAN, which incorporates Wasserstein distance to generate higher-resolution 28x28 images without dimensionality reduction. This advancement represents a significant breakthrough in the field of quantum image generation and establishes a foundation for the application of QGANs in more complex image generation scenarios.

\subsection{Unsupervised I2I translation}
Unsupervised I2I translation aims to translate images from the source domain to the target domain without paired training data while maintaining content consistency, including structural consistency. 
Before CycleGAN \cite{zhu2017unpaired}, several methods mainly relied on prior knowledge \cite{isola2017image} or shared weights \cite{liu2016coupled} to relate the two domains for addressing the challenges of the unpaired setting.
CycleGAN provides a generalized solution for achieving two-sided unsupervised I2I translation by utilizing bidirectional cycle consistency constraints, without depending on task-specific prior knowledge. 
The underlying intuition behind the cycle consistency constraints is that when an image is translated from one domain to the other and then back again, it should arrive where it started.

However, relying solely on bidirectional cycle consistency constraints may lead to loose restrictions and content distortion.
To address this problem, researchers proposed feature-level losses \cite{liu2016coupled,tang2021attentiongan}, which extract feature maps using convolutional neural networks (CNNs) and evaluate the content distance in the deep feature space.
QGAN{\_C} \cite{chen2019quality}introduces image quality evaluation measures such as perceptual loss, allowing for direct comparison of the pixel-wise similarity between the original and reconstructed images rather than comparing feature similarity. Recent research has shown that one-sided unsupervised I2I translation can be achieved using unidirectional cycle consistency constraints, thereby simplifying the architecture and training process compared with two-sided unsupervised I2I translation.
DistanceGAN \cite{benaim2017one} and GCGAN \cite{fu2019geometry} introduce an implicit distance in one-sided unsupervised I2I translation, showcasing competitive translation performance across various applications.

Invertible neural networks (INNs) \cite{dinh2014nice,DBLP:journals/corr/abs-1802-07088,kingma2018glow,dinh2016density} have emerged as a crucial tool for developing reversible methods in unsupervised image translation due to their reversible nature. However, INNs require high computational overhead and are often integrated as a component of the overall model framework,as exemplified by iFlow \cite{dai2021iflowgan} and RevGAN \cite{van2019reversible}.The reversible component specifically operates on data that have already undergone dimensionality reduction during the initial phases of the framework. There is another reversible approach \cite{shen2019towards}where a fully self-reversible generator can be achieved by augmenting the training data through swapping inputs and outputs during the training process, while also implementing separate cycle consistency loss for each mapping direction.
These reversible I2I methods can greatly decrease the number of parameters.
\section{iHQGAN}
\subsection{Problem Formulation}
\label{Problem Formulation}
 We redefine the concept of I2I translation from a quantum perspective,  aiming to identify a forward quantum mapping $\mathcal{F}$: $X \rightarrow Y$ and a corresponding backward quantum mapping $\mathcal{B}$: $Y\rightarrow X$,  such that $\mathcal{B} \circ \mathcal{F} =  \mathcal{F} \circ \mathcal{B}=\mathcal{I}$.
 Given paired training samples $x=\left\{x_i\right\}_{i=1}^N$, $y=\left\{y_i\right\}_{j=1}^N$ ,where  $x \sim p_{\text {data }}(x)$, $y \sim p_{\text {data }}(y)$. The distributions generated by $G$ and $F$ are denoted $p_G$ and $p_F$, respectively. The quantum states $|\phi\rangle=\left\{|\phi\rangle_i\right\}_{i=1}^N$ and
 $|\psi\rangle=\left\{|\psi\rangle_j\right\}_{j=1}^N$ corresponding to the training samples $x$ and  $y$ , respectively. 

\subsection{Overview}
\label{Overview}

\begin{figure}[t]
    \centering
   \includegraphics[width=0.8\textwidth, trim=0cm 0.1cm 0cm 0cm, clip]{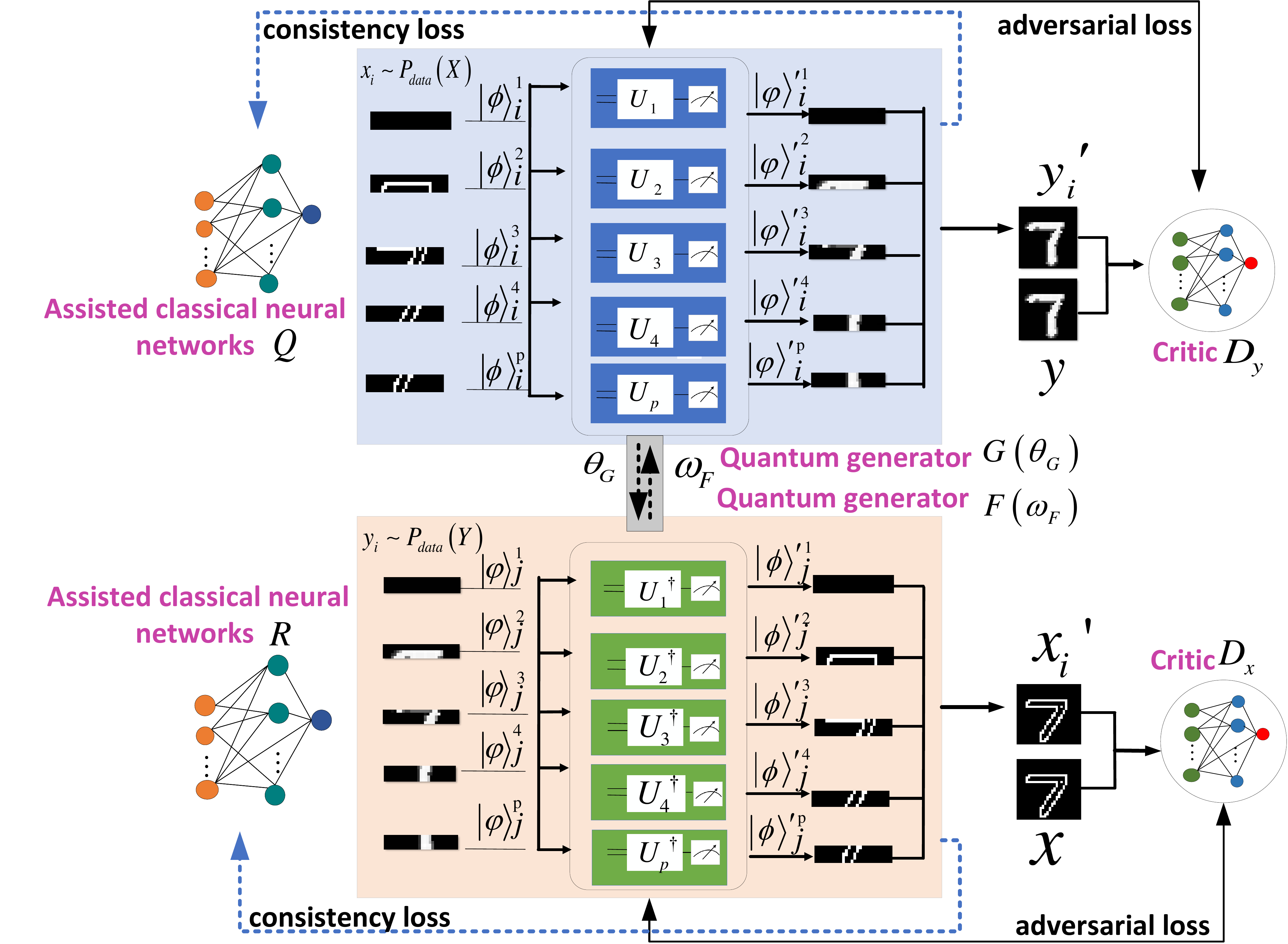}
    \caption{\footnotesize 
    The overall architecture of iHQGAN. iHQGAN consists of two quantum generators, two assisted classical networks(ACNNs), and two classical critics. The quantum generators $G$ and $F$ comprise $p$ sub-quantum generators. ACNNs are utilized to achieve consistency loss, while the critics implement adversarial loss. 
   During the alternating training of the two quantum generators, their respective parameters $\theta_G$ and  $\omega_F$ are interchangeably assigned to facilitate parameter sharing.
    }
    \label{iHQGANmodel}
\end{figure}

Leveraging the inherent approximate reversibility of I2I translation, we propose a groundbreaking reversible hybrid quantum-classical model, iHQGAN, which capitalizes on the unique invertibility of quantum computing, as illustrated in Fig. \ref{iHQGANmodel}. iHQGAN builds upon the PQWGAN \cite{tsang2023hybrid} framework and introduces two key mechanisms: (1) by harnessing the invertibility of quantum computing, we have designed two mutually approximately reversible quantum generators with shared parameters, thereby significantly reducing the parameter scale; (2) we incorporate ACNNs to implement a unidirectional cycle consistency constraint between each quantum generator and its corresponding ACNN, ensuring content consistency between input and output images. Ultimately, through the synergistic interplay of these mechanisms, iHQGAN effectively achieves unsupervised I2I translation.
 
\subsection{Quantum Generator}
Leveraging the inherent invertibility of quantum computing, we have developed two quantum generators, denoted by $G$ and $F$. The parameters associated with these generators are represented by  $\theta_G$ and $\omega_F$, respectively.
Each quantum generator is implemented with a set of quantum circuits. For each quantum circuit in $G$,there is a corresponding quantum circuit in $F$.
To construct a reversible quantum circuit, we need to negate the parameter values of the quantum gates in the original circuit and subsequently arrange all quantum gates in reverse order.Consequently, two mutually reversible quantum circuits are able to share a common set of parameters.

iHQGAN is built upon the PQWGAN framework \cite{tsang2023hybrid}, which employs a patch-based strategy for image generation.
Specifically, the generators $G$ and $F$ are  composed of p small circuits, denoted by $u_k$,  $u_l$ $(k,l=1...p)$.The parameters of these circuits are represented as $({\theta_G})_{k}$ and $({\omega_F})_{l}$, respectively. 
When $k = l$, the circuits satisfy the relationship $u_k u_l = u_l u_k = I$, indicating they are mutual inverses, with 
 $u_l$ being the conjugate transpose of $u_k$,  expressed as 
$u_k = u_l^\dagger$.
To derive $u_l$, we first negate the parameter values of the quantum gates in $u_k$ and then rearrange these quantum gates in reverse order. 
Each quantum sub-circuit consists of $N(N=1,2,..n)$ qubits and includes an amplitude encoding module, a measurement module , and
 $S(S=1,2,..f)$ repeated parameterized blocks. Each 
 block consists of a rotation layer and an entanglement layer. 
The rotation layer is formed by an assembly of single-qubit rotation gates $R$, and the entanglement layer is formed by an assembly of $CNOT$ gates.
Only $R$ gates have  parameters, with three parameters per gate.

\begin{figure}[t]
    \centering
   \includegraphics[width=1\textwidth, trim=0cm 0.1cm 0cm 0cm, clip]{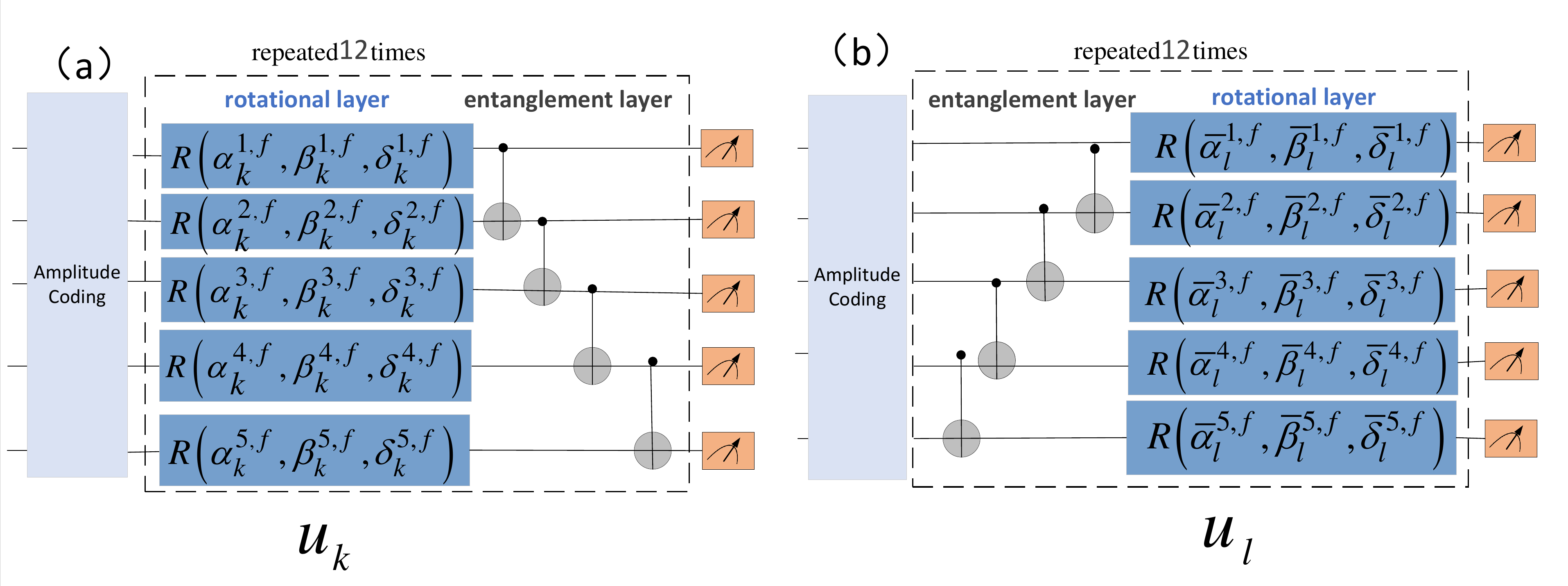}
    \caption{\footnotesize The structure of the two types of quantum circuits. Subfigure(a) depicts the $k-th$ quantum sub-circuit of quantum generator $G$. Subfigure(b) presents the $l-th$ quantum sub-circuit of quantum generator $F$.
    Each quantum sub-circuit has $N = 5$ qubits and $S = 12$ blocks. When $k=l$, the relationship $ u_l $=$ {u_k}^{\dagger} $ holds.
The correspondence of the parameters between the two quantum circuits can be expressed as
$\overline{\alpha}_{k}^{n,f}=\overline{\alpha}_{l}^{n,S-f+1}$,$ \overline{\beta}_{k}^{n,f}=\overline{\beta}_{l}^{n,S-f+1}$,and
$ \overline{\sigma}_{k}^{n,f}=\overline{\sigma}_{l}^{n,S-f+1}$
    }
    \label{iHQGANmodel_circuit}
\end{figure}

We continue to standardize and detail the notation for every parameter of the gates within each block.
The set of parameters for the $k-th$ sub-circuit $u_k$ is denoted by $\theta_{G}^k$.
For $u_k$, the parameters of the quantum gate for the $n-th$ qubit at $f-th$ block include $ \alpha_{k}^{f,n} $, $ \beta_{k}^{f,n} $, and $ \sigma_{k}^{f,n} $.
Similarly, $ \omega_{G}^l$ denotes the set of parameters for the $l-th$ sub-circuit $u_l$. For $u_l$, the quantum gate parameters for the $n-th$ qubit  at $f-th$ block include $ \overline{\alpha}_{l}^{n,f} $, $ \overline{\beta}_{l}^{n,f} $, and $\overline{\sigma}_{l}^{n,f} $.When $k=l$ , we have $ u_l $=$ {u_k}\dagger $.
The correspondence of the parameters between the two subcircuits can be expressed as
$ \overline{\alpha}_{k}^{n,f}=\overline{\alpha}_{l}^{n,S-f+1}$,$ \overline{\beta}_{k}^{n,f}=\overline{\beta}_{l}^{n,S-f+1}$,and
$ \overline{\sigma}_{k}^{n,f}=\overline{\sigma}_{l}^{n,S-f+1}$.
In this study, the quantum generators $G$ and $F$ each consist of $p=32$ sub-circuits. Each quantum sub-circuit has $N = 5$ qubits and $S = 12$  parameterized blocks.
Fig.\ref{iHQGANmodel_circuit} illustrates the structures of the two types of quantum circuits $ u_k $ and $ u_l $.More details of the quantum circuits are provided in \ref{appendixA}.
Table.\ref{tab_generator} presents the common details of two quantum generators $G$ and $F$.

\begin{table}[H]
	\caption{ The common details of two quantum generators $G$ and $F$}
        \label{tab_generator}
        \vspace{-0.2cm} 
	\centering
	\scriptsize
	\begin{threeparttable}
        \setlength{\tabcolsep}{0.6pt} %
		\begin{tabular}{*{5}{>{\centering\arraybackslash}p{2.4cm}}}
			\toprule
			Number of circuit & Qubits of per circuit & Layers of per circuit & Gates of per circuit& Parameter count of per circuit \\[-0.1cm]
		  \midrule
           32& 5 & 24 &60 &   5760\\
             \bottomrule
		\end{tabular}
	\end{threeparttable}
        \vspace{-0.2cm} 
\end{table}

Last, we introduce the process of image translation using the quantum generators $G$ and $F$.
The source images $ x_i $ and $ y_j $ are segmented into patches $ { x_{i}^1, \ldots, x_{i}^p } $ and $ { y_{j}^1, \ldots, y_{j}^p } $,respectively.
These patches are then fed into the corresponding quantum circuits of the quantum generators  $G$ and $F$.
The initial quantum states $ \{ | \phi \rangle_i^1, \ldots, | \phi \rangle_i^p \} $ and $ \{ | \varphi \rangle_j^1, \ldots, | \varphi \rangle_j^p \} $ 
are prepared by encoding the pixel values of each image patch onto $N$ qubits using the amplitude encoding component of each quantum sub-circuit. 
Subsequently, the initial state $\left\{|\phi\rangle_i^1, \cdots,|\phi\rangle_i^p\right\}$ undergoes unitary evolution to yield a new quantum state $|\varphi\rangle \prime _i=\left\{|\varphi\rangle \prime _i^1, \cdots,|\varphi\rangle \prime _i^p\right\}$. Similarly, $|\varphi\rangle_j=\left\{|\varphi\rangle_j^1, \cdots,|\varphi\rangle_j^p\right\}$ evolves unitarily to produce the output  quantum state $|\phi\rangle \prime _j=\left\{|\phi\rangle \prime _j^1, \cdots,|\phi\rangle \prime _j^5\right\}$. 
Following the unitary evolution in each sub-quantum circuit, quantum measurements are performed on the output quantum states to extract classical pixel values 
$\left\{x_i\prime^1,...x_i^p\prime\right\}$ and  $\left\{y_i\prime^1,...y_i\prime^p\right\}$.
Ultimately, these pixel values are reassembled to construct the complete translated images $x_i\prime$ and $y_j\prime$.

\subsection{Unidirectional cycle consistency constraint 
} 
\begin{figure}[t]
    \centering
   \includegraphics[width=0.8\textwidth, trim=0cm 0.1cm 0cm 0cm, clip]{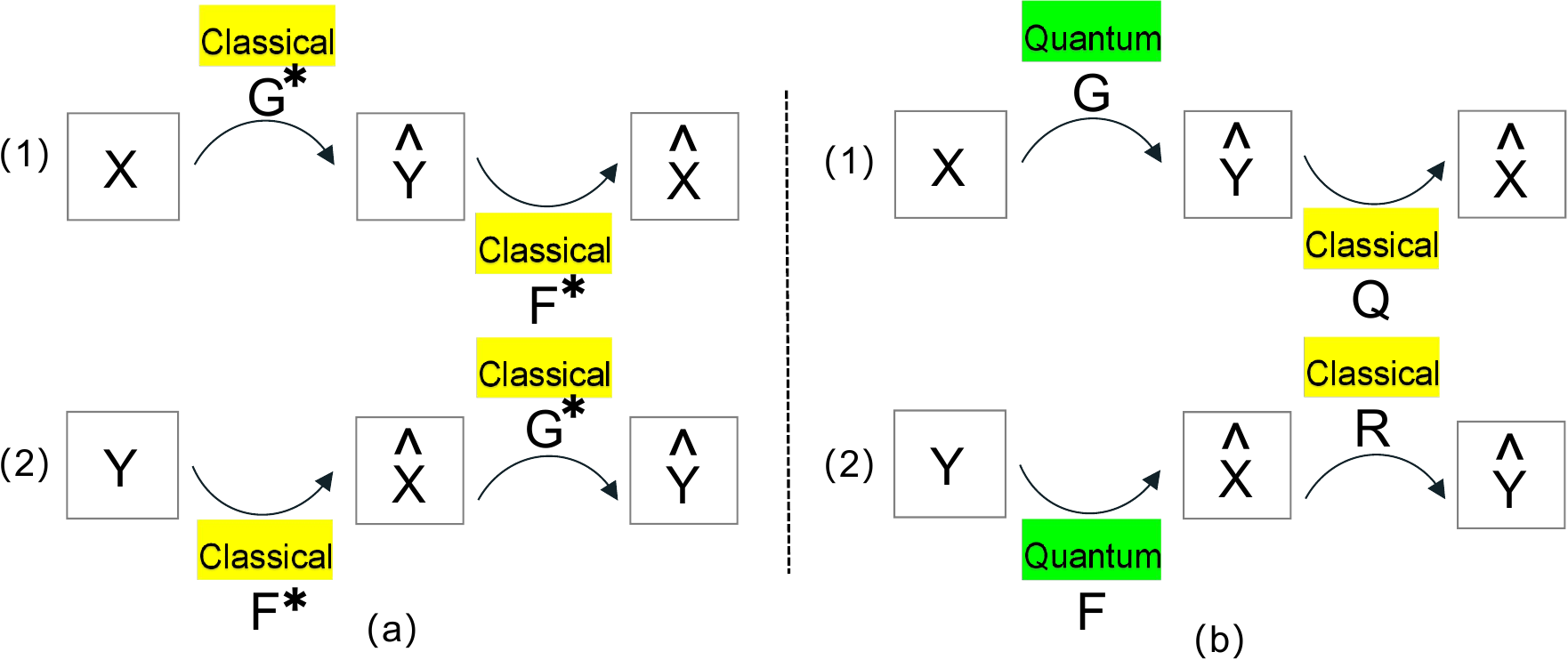}
    \caption{\footnotesize  Two ways of implementing cycle-consistent constraints. 
    \textbf{Left}: bidirectional cycle consistency constraints between two classical generators.Sub-figure(1) shows forward cycle-consistent constraints $ x \rightarrow G^{*}(x) \rightarrow F^{*}(G^{*}(x)) \approx x. $ Sub-figure(2) shows backward cycle consistency constraints  $ x \rightarrow G^{*}(y) \rightarrow F^{*}(G^{*}(y)) \approx y $ \textbf{Right}: The unidirectional constraints between each quantum generator and its corresponding ACNN. Sub-figure(1) shows forward cycle consistency constraints $ x \rightarrow G(x) \rightarrow Q(G(x)) \approx x.$ Sub-figure(2) shows forward cycle consistency constraints  $ x \rightarrow F(y) \rightarrow R(F(y)) \approx y $
    }
    \label{Mapping}
\end{figure}

In the realm of classical unsupervised I2I  translation, researchers proposed the strategy of unidirectional cycle consistency constraints between two generators, with one acting as the primary generator and the other as the auxiliary generator \cite{fu2019geometry,benaim2017one}. Additionally, numerous studies have demonstrated that classical components can enhance the performance of quantum circuits \cite{callison2022hybrid,chang2024latent,shu2024variational}. Building on the above two points, we introduced the strategy of unidirectional cycle consistency constraint to iHQGAN, implementing it between each quantum generator and its corresponding assisted classical neural networks (ACNN), where the ACNN helps optimize the parameter space of the quantum generator. 
While bidirectional cycle consistency constraints are commonly applied between two generators in the classical domain,  
the application of such constraints between two quantum generators may hinder model performance due to the limited expressiveness of quantum generators.
Fig.\ref{Mapping} illustrates two different ways to implement cycle consistency constraints. 
The related discussion is detailed in Section \ref{Analysis of the unidirectional cycle consistency constraint} of this paper.

 Each ACNN comprises six layers, including both an encoding block and a decoding block. 
The four encoding layers and the first layer of the decoding block utilize a LeakyReLU activation function with a slope of 0.05, while the last layer of the decoding block employs a Tanh activation function.

\subsection{Loss Function}
\textbf{Adversarial Loss} 
The purpose of the adversarial loss function \cite{goodfellow2020generative} is to minimize the difference between the distribution of the learned data and that of the real data .
For the mapping function $\mathcal{F}$: $X \rightarrow Y$ and its corresponding  critic  $D_Y$, the objective ${L}_{{GAN}}^{{D_Y}}\left({G}, {D}_{{Y}}, {x}, {y}\right)$ is formulated as:
\begin{equation}
\vspace{-0.8cm}
\small
\begin{aligned}
{L}_{{GAN}}^{{D_Y}}\left({G}, {D}_{{Y}}, {x}, {y}\right)=
\mathbb{E}_{y \sim p_{\text {data }}(y)}
[D_Y(\boldsymbol{y})]-\\
\mathbb{E}_{y \sim p_{\text {data }}(y)}[D_Y(G(\boldsymbol{x}))]- 
\lambda \mathbb{E}_{\hat{\boldsymbol{y}} \sim P_{\text {data }}{\hat{\boldsymbol{(y)}}}}\left[\left(\left\|\nabla_{\hat{\boldsymbol{y}}} D_Y(\hat{\boldsymbol{y}})\right\|_2-1\right)^2\right]
\end{aligned}
\vspace{0.8cm}
\label{Loss:Dy}
\end{equation}
where $\hat{\boldsymbol{y}}$ is a distribution uniformly sampled between $y\sim p_{\text {data }}(y)$ and $y\sim p_G$  and $\lambda$ is the gradient penalty coefficient.
A similar loss for the backward quantum mapping $\mathcal{B}$: $Y \rightarrow X$ and its critic  $D_X$ is denoted by $\scriptsize{{L}_{{GAN}}^{{{D_Y}}}\left({F}, {D}_{{X}}, {x}, {y}\right)}$.
\begin{equation}
\vspace{-0.8cm}
\small
\begin{aligned}
{L}_{{GAN}}^{{D_X}}\left({G}, {D_X}_{{Y}}, {x}, {y}\right)=
\mathbb{E}_{x \sim p_{\text {data }}(x)}
[D_X(\boldsymbol{x})]-\\
\mathbb{E}_{x \sim p_{\text {data }}(x)}[D_X(F(\boldsymbol{x}))]- 
\lambda \mathbb{E}_{\hat{\boldsymbol{x}} \sim P_{\text {data }}{\hat{\boldsymbol{(x)}}}}\left[\left(\left\|\nabla_{\hat{\boldsymbol{x}}} D_X(\hat{\boldsymbol{x}})\right\|_2-1\right)^2\right]
\end{aligned}
\vspace{0.8cm}
\label{Loss:Dx}
\end{equation}
Further details on the critic can be found in \ref{appendixB}.

\textbf{Cycle Consistency Loss}
Utilizing the cycle consistency loss \cite{zhu2017unpaired}, we implemented a unidirectional cycle consistency constraint between each quantum generator and its corresponding ACNN to maintain content consistency.
\begin{equation}
{L}^G_{\text {cyc }}(G, Q)=\mathbb{E}_{x \sim p_{\text {data }}(x)}\left[\|Q(G(x))-x\|_1\right]
\end{equation}
\begin{equation}
{L}^F_{\text {cyc }}(G, F)=\mathbb{E}_{y \sim p_{\text {data }}(y)}\left[\|R(F(y))-y\|_1\right]
\end{equation}

\textbf{Image Quality-aware (IQA) Loss }  
To generate higher fidelity images, Chen et al. \cite{chen2019quality} introduced a quality-aware loss that compares the source and reconstructed images at the domain level. We chose the structural similarity index (SSIM) metric for the image quality-aware loss to effectively maintain structural consistency between the input and output images.
\begin{equation}
{L}^G_{\text {IQA}}(G, Q)=1-SSIM(Q(G(x)),x)
\end{equation}
\begin{equation}
{L}^F_{\text {IQA}}(F, R)=1-SSIM(R(F(y)),y)
\end{equation}
The complete objective function for each quantum generator is formulated as follows:
\begin{equation}
\mathrm{L^G}=\varepsilon * \mathrm{~L}^G_{\mathrm{GAN}}+\eta * \mathrm{~L}^G_{c y c}+\rho * \mathrm{~L}^G_{IQA}
\label{Loss:G}
\end{equation}

\begin{equation}
\mathrm{L^F}=\varepsilon * \mathrm{~L}^F_{\mathrm{GAN}}+\eta * \mathrm{~L}^F_{c y c}+\rho * \mathrm{~L}^F_{IQA}
\label{Loss:F}
\end{equation}

Where $\varepsilon$, $\eta$, and $\rho$  are the weight coefficients corresponding to the three components.

\subsubsection{Training and optimization} 
We implemented an alternating training strategy that involves mutually assigning weights between the quantum generators $G$ and $F$. During the training process, for each training batch, we first train $G$ and update its parameters, then pass the updated parameters of $G$ to $F$. Next, we train $F$ and update its parameters, and then pass the updated parameters of $F$ back to $G$ for further training. This process is repeated continuously until the model reaches a Nash equilibrium. The complete training algorithm for iHQGAN is detailed in Algorithm \ref{alg}.

\begin{breakablealgorithm}
    \renewcommand{\algorithmicrequire}{\textbf{Input:}}
    \caption{Pseudocode of training algorithm.}
    \label{alg}
    \begin{algorithmic}[1]
        \Require Gradient penalty coefficient $\lambda$,  weight coefficients in the loss function per generator $\epsilon$,$\eta$,$\rho$, critic iterations per generator iteration batch size $n_c$, number of epochs $n_{\text{epochs}}$, batch size $m$, Adam hyperparameters $\eta_1$, $\eta_2$, $\beta_1$, $\beta_2$.random number $\xi \sim U[0, 1]$   
        \State Initialize critic parameters $\kappa_{D_y}$, 
        $\kappa_{D_x}$, generator parameters $\theta_{G}$, $\omega_{F}$. $\theta_{G}$ and $\omega_{F}$ contain  parameters $\alpha$, $\beta$, $\delta$ and $\alpha^{*}$, $\beta^{*}$, $\delta^{*}$ respectively.
        \State $\alpha^{*}$, $\beta^{*}$, $\delta^{*} := \alpha$, $\beta$, $\delta$
        \State $batchnum:=0$
        \For{$\text{epoch} = 1, \ldots, n_{\text{epochs}}$}
            \For{$i = 1, \ldots, m$}
                 \State Sample real data $x \sim P_{\text{data}}(x)$, $y \sim P_{\text{data}}(y)$
                 \State Encoding $x$ into amplitude  of $\left| \phi \right\rangle$
                  \State $y^{\prime} \stackrel{\text{decoding}}{\longleftarrow}\left| \psi \right\rangle^{\prime} \leftarrow G(\theta_{G}, \left| \phi \right\rangle)$ 
                  \State $\hat{y} \leftarrow \xi y + (1 - \xi) y^{\prime}$
                  \State $ {L_{GAN}^{{D_Y}(i)}}\leftarrow D_{Y}({y^{\prime}}) - D_{Y}(y) + \lambda \left( \left\| \nabla_{\hat{y}} D_{Y}(\hat{y}) \right\|_2 - 1 \right)^2$ 
                  \State $\kappa_{D_y} \leftarrow \text{Adam} \left( \frac{1}{m} \sum_{i=1}^{m} {L_{GAN}^{{D_Y}(i)}},\kappa_{D_y}, \eta_1, \beta_1, \beta_2 \right)$ 
                 \State Encoding $y$ into amplitude  of $\left| \psi \right\rangle$
                  \State $x^{\prime} \stackrel{\text{decoding}}{\longleftarrow}\left| \psi \right\rangle^{\prime} \leftarrow F(\omega_{F}, \left| \psi \right\rangle)$ 
                  \State $\hat{x} \leftarrow \xi x + (1 - \xi) x^{\prime}$
                  \State ${L_{GAN}^{{D_X}(i)}}\leftarrow D_{X}({x^{\prime}}) - D_{X}(x) + \lambda \left( \left\| \nabla_{\tilde{x}} D_{X}(\hat{x}) \right\|_2 - 1 \right)^2$        
                  \State $\kappa_{D_x} \leftarrow 
                  \text{Adam} \left( \frac{1}{m} \sum_{i=1}^{m}{L_{GAN}^{{D_X}(i)}}, \kappa_{D_x}, \eta_2, \beta_1, \beta_2 \right)$
                \State $batchnum:=batchnum+1$ 
                \If{$batchnum/n_c==0$} 
                    \State Encoding $x$ into amplitude  of $\left| \phi \right\rangle$
                    \State $y^{\prime} \stackrel{\text{decoding}}{\longleftarrow}\left| \psi \right\rangle^{\prime} \leftarrow G(\theta_{G}, \left| \phi \right\rangle)$ 
                    \State $\hat{y} \leftarrow \xi y + (1 - \xi) y^{\prime}$
                    \State $ L^{G(i)}\leftarrow \epsilon(D_{Y}({y^{\prime}}) - D_{Y}(y) + \lambda \left( \left\| \nabla_{\hat{y}} D_{Y}(\hat{y}) \right\|_2 - 1 \right)^2)+\eta\Vert(Q(G(x))-y) \Vert_1+\rho 
                    (1-SSIM(Q(G(x),x))$
                    \State $\theta_{G} \leftarrow \text{Adam} \left( \frac{1}{m} \sum_{i=1}^{m} L^{G(i)}, \theta_{G}, \eta_1, \beta_1, \beta_2 \right)$
                    \State $\alpha^{*}$, $\beta^{*}$, $\delta^{*} := \alpha$, $\beta$, $\delta$
                
                    \State Encoding $y$ into amplitude  of $\left| \psi\right\rangle$ 
                    \State $x^{\prime} \stackrel{\text{decoding}}{\longleftarrow}\left| \phi \right\rangle^{\prime} \leftarrow F(\omega_{G}, \left| \psi\right\rangle)$   
                    \State $\hat{x} \leftarrow \xi x + (1 - \xi) x^{\prime}$
                    \State $ L^{F(i)}\leftarrow\epsilon( D_{X}({x^{\prime}}) - D_{X}(x) + \lambda \left( \left\| \nabla_{\hat{x}} D_{X}(\hat{x}) \right\|_2 - 1 \right)^2)+\eta\Vert(R(F(Y))-x) \Vert_1+\rho
                    (1-SSIM(R(F(y),y))$
                    \State $\theta_{F} \leftarrow \text{Adam} \left( \frac{1}{m} \sum_{i=1}^{m} L^{F(i)}, \theta_{F}, \eta_2, \beta_1, \beta_2 \right)$
                    
                    \State $\alpha$, $\beta$, $\delta$:=$\alpha^{*}$, $\beta^{*}$, $\delta^{*}$
               \EndIf
            \EndFor
        \EndFor
   \end{algorithmic}
\end{breakablealgorithm}

\subsubsection{Post-processing images generated by iHQGAN}
\begin{figure}[H]
    \centering
  \includegraphics[width=0.35\textwidth, trim=0cm 0.0cm 0cm 0cm, clip]{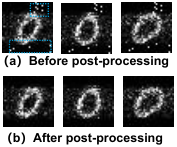}
    \caption{\footnotesize 
    Post-processing of images generated by iHQGAN. (a) Images before post-processing; (b) Images after post-processing.
    }
    \label{PostProcessing}
\end{figure}

The images generated by PQWGAN \cite{tsang2023hybrid} often exhibit special discrete noise. This issue is also evident in iHQGAN, as illustrated in the light blue dashed box of Fig.\ref{PostProcessing}(a).
To improve  image quality, we performed simple post-processing on the generated images without significantly affecting the features of the images directly output by iHQGAN. 
Specifically, we set the pixel values to zero for rows \( i \) in the ranges \( [0, 7] \) and \( [26, 31] \).
The denoising results are shown in Fig.\ref{PostProcessing}(b).

\section{Experiments}
\label{Experiments }
This section presents a rigorous experimental evaluation of the iHQGAN model, providing comprehensive empirical evidence to demonstrate its effectiveness, generalizability , and parameter-saving nature.
\subsection{Datasets and Evaluation Metrics}
\label{Datasets and Evaluation Metric}
\begin{figure}[H]
	\centering
	\includegraphics[width=0.9\textwidth]{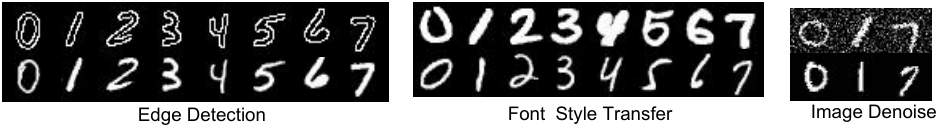}
	\caption{
		\footnotesize
            Examples of the datasets. 
            The datasets consist of three I2I translation tasks with 19 sub-datasets: 8 for \textit{Edge Detection}(labels 0-7), 8 for \textit{Font Style Transfer} (labels 0-7), and 3 for \textit{Image Denoising }(labels 0, 1, 7).
		}
       \label{trainDataset}
\end{figure}

 \textbf{Datasets} The current state-of-the-art QGANs are limited by quantum resources. They are suited for training on small grayscale image datasets, such as MNIST\cite{lecun-mnisthandwrittendigit-2010} and MNIST-C\cite{DBLP:journals/corr/abs-1906-02337}. We created three unpaired I2I translation tasks: \textit {Edge Detection}, \textit{Font Style Transfer}, and \textit{Image Denoising} using MNIST datasets, resulting in a total of 19 sub-datasets.
 Each sub-dataset comprises 1,250 images, with the training and test sets split in an 8:2 ratio. Fig.\ref{trainDataset} presents some examples from the datasets. Table.~\ref{tab_dataset} provides the basic details of datasets.  Additional details on the datasets can be found in \ref{appendixC}.
 To facilitate amplitude coding for images using a 5-qubit quantum circuit, we preprocess the original 28×28 images by padding them to 32×32.

\begin{table}[H]
	\caption{Datasets with basic setup used}
	\label{tab_dataset}
        \vspace{-0.2cm} 
	\centering
	\scriptsize
	\begin{threeparttable}
        \setlength{\tabcolsep}{0.5pt} %
		\begin{tabular}{*{5}{>{\centering\arraybackslash}p{2.6cm}}}
			\toprule
			Dataset & No.of training Instance & No.of test Instance
   & Size of Instance
   & No.of Classes \\[-0.1cm]
			\midrule
			\textit{Edge Detection} & $1000$ &$250$
   &28$\times$28
   &$8$ object categories   \\
			\textit{Font  Style Transfer} & $1000$& $250$
    &28$\times$28
   &$8$ object categories  \\
			\textit{Image Denoising} & $1000$& $250$
    &28$\times$28
&$3$ object categories \\
			\bottomrule
		\end{tabular}
	\end{threeparttable}
        \vspace{-0.2cm} 
\end{table}
 \textbf{Evaluation Metric} For quantitative evaluation, we utilize  Fréchet inception distance (FID), structural similarity index (SSIM), and peak signal-to-noise ratio(PSNR) to assess the performance of the methods, as well as to investigate the number of parameters.

\begin{enumerate}[label=(\roman*)]
  \item FID \cite{silver2023mosaiq} measures the distance between the distributions of the generated samples and the real samples. A lower score indicates that the model can produce higher-quality images.The FID formula is:
\begin{equation}
F I D=\left\|\mu_r-\mu_g\right\|^2+\operatorname{Tr}\left(\Sigma_r+\Sigma_g-2\left(\Sigma_r \Sigma_g\right)^{1 / 2}\right)
\end{equation}
where $\mu_{r}$ and $\mu_{g}$ represent the means of the features for real and generated images, respectively.$\Sigma_{r}$ and $\Sigma_{g}$ are the feature covariance matrices.$Tr$ denotes the matrix trace.
  \item SSIM \cite{amirkolaee2022development} evaluates the structural similarity between two images, with higher values reflecting greater similarity and aligning more closely with human perception of image quality.The SSIM formula is:
 \begin{equation}
\operatorname{SSIM}=\frac{\left(2 \mu_x \mu_y+c_1\right)\left(2 \sigma_{x y}+c_2\right)}{\left(\mu_x^2+\mu_ y^2+c_1\right)\left(\sigma_x^2+\sigma_y^2+c_2\right)}
\end{equation}
where,\( \mu_x \) and \( \mu_y \) denote the mean values of images \( x \) and \( y \),
\( \sigma_x^2 \) and \( \sigma_y^2 \) are the variances of images , and  \( \sigma_{xy} \) represents their covariance.$c_1=(0.01 L)^2$ and $c_2=(0.03 L)^2$ are regularization constants.

  \item PSNR \cite{kancharagunta2019csgan} is a common metric for assessing the difference between original and reconstructed images,  focusing primarily on pixel-level discrepancies.
  A higher PSNR value indicates less distortion and better quality of the generated samples.The PSNR formula is as follows:
\begin{equation}
P S N R=10 \cdot \log _{10}\left(\frac{R^2}{MS E}\right)
\end{equation}
where $R$ is the maximum  pixel value in an image and $MSE$ is the mean squared error between two images.

\end{enumerate}
\subsection{Implementation Details}
 The experiment was simulated using PyTorch, and PennyLane on a 13th Gen Intel(R) Core(TM) i5-13490F CPU with 32.0GB of RAM. 
 We utilized the Adam optimizer with initial learning rates of 0.0002 for classical critics  and ACNNs, and 0.01 for quantum generators. The two weight decay hyperparameters of the Adam optimizer were set to 0 and 0.9. The model underwent training for 50 epochs with a batch size of 10.  The hyperparameters for the loss function of each discriminator, as defined in Equations \ref{Loss:Dy} and \ref{Loss:Dx} , were  set to $\lambda = 10$ and 
 the hyperparameters of the loss function of each quantum generator, as defined in Equations \ref{Loss:G} and \ref{Loss:F}, were  set to $\varepsilon  = 10, \eta = 20$, and $\rho = 300$.
\subsection{Baselines}

\textit{Edge Detection} and \textit{Font Style Transfer}  are utilized to evaluate iHQGAN’s effectiveness and demonstrate that iHQGAN can reduce the number of parameters in classical  irreversible methods through a reversible mechanism, similar to classical reversible methods. We selected the classical irreversible method CycleGAN \cite{zhu2017unpaired} and the reversible method One2One \cite{shen2020one} as baselines.\textit{Image Denoising} is used to further assess the generalizability of iHQGAN, with the classical nonlinear,CNN-based CycleGAN and the classical linear Gaussian filter serving as baselines.

We implement the CycleGAN and One2One using low-complexity CNN generators  
to preliminarily validate
the aforementioned objective.
Further details on the baselines are provided below. The generator configurations of the baselines are summarized in Table~\ref{tab:generator_configurations}. For the Gaussian filter, we utilized the \texttt{cv2.GaussianBlur} function from the OpenCV library\cite{opencv_library}, with a 5×5 Gaussian kernel size defining the neighborhood range, and the standard deviation set to 0.

\begin{table}[H]
    \scriptsize
    \centering
    \caption{The generator configurations of baselines}
    \label{tab:generator_configurations}
    \begin{threeparttable}
         \begin{tabular}{|>{\centering\arraybackslash}p{2cm}|>{\centering\arraybackslash}p{4cm}|>{\centering\arraybackslash}p{6cm}|}
            \hline
            \textbf{Model} & \textbf{Task} & \textbf{Generator configurations (\tnote{1} $Ck$-\tnote{2} $Dm$)}  \\ \hline
            CycleGAN & Font Style Transfer & $C12-D1$  \\ \cline{2-3}
                      & Edge Detection      & $C70-D1$ \\ \hline
            One2One  & Font Style Transfer & $C12-D1$ \\ \cline{2-3}
                      & Edge Detection     & $C70-D1$  \\ \hline
        \end{tabular}
        \begin{tablenotes}
            \scriptsize
            \item[1] $Ck$ denotes a 16 × 16 Convolution-LeakyReLU layer with \( k \) channels, stride 16, and no padding. 
            \item[2]  $Dm$  represents a 16 × 16 Deconvolution-LeakyReLU layer with \( m \) channels, stride 16, and no padding.
        \end{tablenotes}
    \end{threeparttable}
\end{table}

\subsection{Result}
\label{Evaluation and Analysis}
We conducted experiments on the \textit{Edge Detection}, \textit{Font Style Transfer} datasets, and \textit{Image Denoising}, yielding both quantitative and qualitative results.  To facilitate comparison, the methods involved are labeled as 'iHQGAN', 'CycleGAN', 'One2One', and 'Gaussian filter'. The images from the source domain are labeled as 'Source', while the ground truth images manually created for comparison are labeled as 'GT'.

\begin{figure}[H]
    \centering
   \includegraphics[width=1\textwidth, trim=0cm 0cm 0cm 0cm, clip]{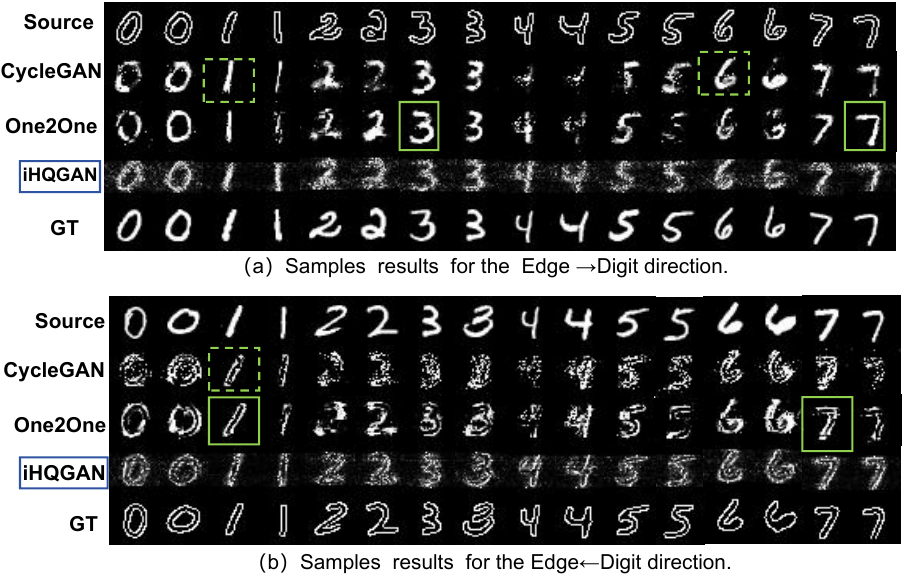}
    \caption{
    Qualitative comparison 
    of outputs generated by various methods
  for the mapping Edge $\leftrightarrow$ Digit trained on the \textit{Edge Detection} datasets. From top to bottom: Source, CycleGAN, One2One, iHQGAN, and GT(ground truth).
  The images generated by iHQGAN effectively achieve domain transfer while maintaining structural consistency with the source images. They are smoother, with no noticeable pixel loss. The green dashed and solid lines highlight that CycleGAN and One2One generate only a few high-quality samples, respectively, while others exhibit pixel loss and fail to preserve structural consistency with the source images.
  The blue solid boxes highlight the key experiment.  
    }
    \label{Edge_Detection}
\end{figure}

  \begin{figure}[H]
    \centering
   \includegraphics[width=1\textwidth, trim=0cm 0cm 0cm 0cm, clip]{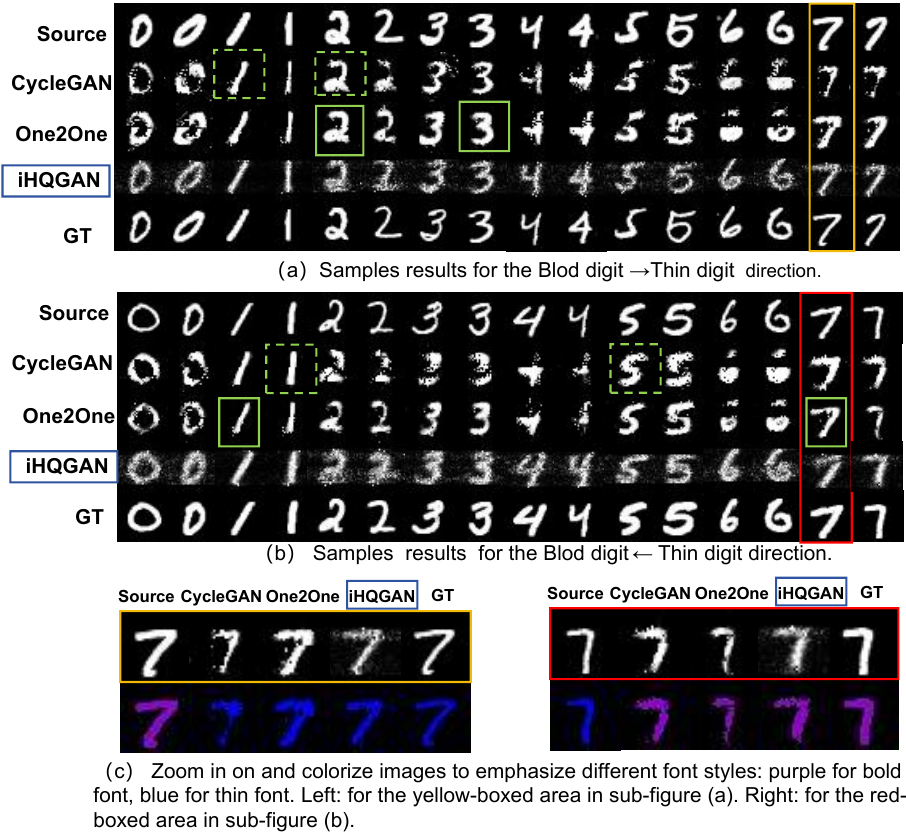}
    \caption{
    Qualitative comparison of outputs generated by various methods trained on \textit{Font Style Transfer} datasets for the mapping Bold Digit $\leftrightarrow$ Thin Digit. 
    From top to bottom: Source, CycleGAN, One2One, iHQGAN, and GT(ground truth).  
   In contrast, the green dashed and solid lines highlight that CycleGAN and One2One respectively generate only a few high-quality samples, while others exhibit pixel loss and fail to maintain structural consistency with the source images. Subfigure (c) emphasizes the poor effectiveness of domain transfer in CycleGAN and One2One, as the font styles of the generated images are not easily distinguishable from those of the source images. 
    The blue solid boxes highlight the key experiment.
    }
    \label{Font_Style_Transfer}
\end{figure}

\subsubsection{Experiment on different unsupervised I2I translation tasks}
\label{Experiment on different unsupervised I2I translation tasks}
\textbf{Edge Detection and Font Style Transfer}
For \textit{Edge Detection} and \textit{Font Style Transfer},  iHQGAN effectively performs domain transfer and maintains structural consistency between the source images and outputs. Moreover, iHQGAN is superior to  CycleGAN and the One2One, as depicted in Fig.\ref{Edge_Detection} and Fig.\ref{Font_Style_Transfer}.
First, the images produced by iHQGAN exhibit more complete and smoother content, while the other two methods suffer from pixel loss issues. This phenomenon is particularly evident in Fig.\ref{Edge_Detection} when compared with Fig.\ref{Font_Style_Transfer}. 
Additionally, iHQGAN maintains better structural consistency across all samples, while CycleGAN and One2One preserve structural consistency only in a few cases, performing poorly overall. The few good results for CycleGAN and One2One are highlighted with green dashed and solid bounding boxes.
For \textit{Font Style Transfer}, we zoom in on and colorize some examples of results to emphasize different font styles, as shown in subfigure (c) of Fig.\ref{Font_Style_Transfer}. Images produced by iHQGAN exhibit more obvious changes in font style than those produced by the other two classical methods.

We conducted quantitative assessments of image quality across different methods using various quality metrics.
In Table.~\ref{Edge_Detection_table}, the optimal results for the 
Edge $\longrightarrow$ Digit
direction 
primarily appear among One2One and iHQGAN, while in the Edge $\longleftarrow$  Digit direction, iHQGAN encompasses nearly all optimal values. Table.~\ref{Font_Style_Transfer_table} indicates that in both the Bold Digit $\longrightarrow$ Thin Digit and Bold Digit $\longleftarrow$  Thin Digit directions, the optimal results are similarly concentrated in iHQGAN. 
This suggests that although classical methods may outperform iHQGAN in certain cases, overall, iHQGAN demonstrates excellent performance in both the \textit{Edge Detection} and \textit{Font Style Transfer} tasks,
showcasing the balanced performance of its two generators.

The effectiveness of the iHQGAN framework arises from the synergistic effects of mechanisms such as the suitable loss function, the unidirectional cycle consistency constraint, two mutually reversible generators with shared parameters, and the alternating training strategy.
Table.~\ref{Parameter_counts_table} reports the parameter counts of various methods used in the \textit{Edge Detection} and \textit{Font Style Transfer}. 
In both tasks, iHQGAN and One2One require only a single generator's parameter set.
In contrast, CycleGAN requires approximately twice as many parameters as iHQGAN and One2One.
This is because both iHQGAN and One2One are reversible, whereas CycleGAN is irreversible.
\begin{table}[H]
\centering
\caption{The parameter count of different methods used for \textit{Edge Detection }and \textit{Font Style Transfer} }
\fontsize{7}{12}\selectfont
\begin{threeparttable}     
\begin{tabular*}{0.6\linewidth}{cccc}
\hline
Task& CycleGAN &One2One &iHQGAN  \\
\hline
Edge Detection &35911 $\times$ 2 &35911&5760\\ 
Font Style Transfer & 6157 $\times$ 2 & 6157 & 5760\\
\hline
\end{tabular*}
\end{threeparttable}
\label{Parameter_counts_table}
\end{table}

\begin{table}[H]
\belowrulesep=0pt
\aboverulesep=0pt
\centering
\caption{Quantitative evaluation of image quality generated by various methods used for \textit{Edge Detection}.}
\fontsize{7}{12}\selectfont
\label{tab_4}
\renewcommand{\arraystretch}{1.5}
\setlength{\tabcolsep}{0.1pt}
\begin{threeparttable}     
\begin{tabular}{|c|ccc|ccc|ccc|ccc|ccc|ccc|ccc|}
\hline
\multirow{3}{*}{Dataset} & \multicolumn{9}{c|}{Edge $\longrightarrow$ Digit} &\multicolumn{9}{c|}{Edge $\longleftarrow$Digit} \\
\cline{2-19}
& \multicolumn{3}{c|}{CycleGAN} &\multicolumn{3}{c|}{One2One} &\multicolumn{3}{c|}{iHQGAN} &\multicolumn{3}{c|}{CycleGAN} & \multicolumn{3}{c|}{One2One} &\multicolumn{3}{c|}{iHQGAN} \\

\cmidrule(lr){2-4}\cmidrule(lr){5-7} \cmidrule(lr){8-10} \cmidrule(lr){11-13} \cmidrule(lr){14-16} \cmidrule(lr){17-19}

&\tnote{1} FID&SSIM&PSNR &FID &\tnote{1} SSIM& PSNR &FID &SSIM &\tnote{1} PSNR &SSIM &FID &PSNR&FID&SSIM&PSNR&FID&SSIM&PSNR \\
\hline
lable0  &31.36 &0.38 &11.43 
        &\tnote{2} \textbf{20.77} &0.41 &10.62
        &22.46 &\textbf{0.59} &\textbf{12.53}
        
        &43.23 &0.31 &9.15 
        &31.62 &0.26 &8.96
        &\textbf{18.95}&\textbf{0.5} &\textbf{11.64}\\

lable1  &\textbf{5.48} &0.65 &16.4
        &8.65 &0.53 &15.06
         &8.12&\textbf{0.72} &\textbf{17.74}
         
         &12.45 &0.39 &13.63
        &13.73 &0.32 &12.87
        &\textbf{11.20}&\textbf{0.51} &\textbf{14.81}\\
        
lable2   &27.55 &0.43 &\textbf{12.06}
        &\textbf{19.46} &0.44 &11.93
        &22.48 &\textbf{0.47} &11.88
        
         & 33.33 & 0.25 & 9.89
        &26.20 & 0.29 &10.24
        &\textbf{19.44}&\textbf{0.33}  &\textbf{11.0} \\

lable3  &16.00 &0.27 &10.05
        &\textbf{13.31} &\textbf{0.57} &\textbf{12.96}
        &19.98&0.30 & 10.92
        
         &33.25 &0.27 &10.05
        &21.96 &0.36 &19.65
        &\textbf{18.70}&\textbf{0.43}&\textbf{11.56} \\
        
lable4  &36.91 &0.25 &12.22
        &25.60 &0.37 &12.36
        &\textbf{18.71} &\textbf{0.49} &\textbf{12.69}
        
         &41.29 &0.24 &10.36
         &28.23&0.27 &10.48
         &\textbf{16.83}&\textbf{0.38}&\textbf{11.80} \\
        
lable5  &24.79 &0.43 &12.31
        &\textbf{17.72} &\textbf{0.44} &\textbf{12.49}
        &21.01 &0.32 &11.03
        
         &34.13 &0.27 &9.99
        &25.18 &0.3 &10.31
        &\textbf{19.35}&\textbf{0.39} &\textbf{11.31}\\
        
lable6 &21.02 &\textbf{0.52} &\textbf{12.46}
        &\textbf{20.32} &0.44 &12.09
        &20.60 &0.48 &12.01
        
         &30.95 &0.28 &10.3
        &25.20 &0.31 &10.42
         &\textbf{17.85}&\textbf{0.42} &\textbf{11.66}\\
        
lable7   &21.84 &0.47 &13.33
        &\textbf{8.95} &\textbf{0.64} &\textbf{14.95}
        &15.36 &0.57 &13.93
        
         &26.22 &0.28 &10.88
        &17.64 &0.41 &11.73
        &\textbf{11.61}&\textbf{0.41} &\textbf{12.35}\\
\hline
\end{tabular}
    \begin{tablenotes} 
        \footnotesize
        \item[1] A lower FID is better, while higher SSIM and PSNR values are more desirable.
        \item[2] Bold text indicates the best-performing values within a set of experiments.
        
      \end{tablenotes} 
\end{threeparttable}
\label{Edge_Detection_table}
\end{table}
\begin{table}[H]
\belowrulesep=0pt
\aboverulesep=0pt
\centering
\caption{Quantitative evaluation of image quality generated by various methods used for \textit{Font Style Transfer}.}
\fontsize{7}{12}\selectfont
\renewcommand{\arraystretch}{1.5}
\setlength{\tabcolsep}{0.1pt}
\begin{threeparttable}     
\begin{tabular}{|c|ccc|ccc|ccc|ccc|ccc|ccc|ccc|}
\hline
\multirow{3}{*}{Dataset} & \multicolumn{9}{c|}{Bold Digit$\longrightarrow$ Thin Digit} &\multicolumn{9}{c|}{Bold Digit $\longleftarrow$ Thin Digit} \\
\cline{2-19}
& \multicolumn{3}{c|}{CycleGAN} &\multicolumn{3}{c|}{One2One} &\multicolumn{3}{c|}{iHQGAN} &\multicolumn{3}{c|}{CycleGAN} & \multicolumn{3}{c|}{One2One} &\multicolumn{3}{c|}{iHQGAN} \\

\cmidrule(lr){2-4}\cmidrule(lr){5-7} \cmidrule(lr){8-10} \cmidrule(lr){11-13} \cmidrule(lr){14-16} \cmidrule(lr){17-19}

&\tnote{1} FID&SSIM&PSNR &FID & \tnote{1} SSIM& PSNR &FID &SSIM &\tnote{1} PSNR &FID&SSIM&PSNR&FID&SSIM&PSNR&FID&SSIM&PSNR \\
\hline
label0&38.67 &0.61&11.57&34.30&0.66 &11.47 &\tnote{2} \textbf{15.27} &\textbf{0.77} &\textbf{14.67} &47.21 &0.64 &10.91&37.52&0.62 &11.27 &\textbf{25.79} & \textbf{0.74} &\textbf{12.78} \\
label1 &7.92 &0.82 &19.53 &10.00 &0.74 &17.08 &\textbf{5.34} &\textbf{0.84} & \textbf{19.81} &9.84 &0.84 &18.56&17.97 &0.63 &15.13 & \textbf{7.00} & \textbf{0.87} & \textbf{18.7} \\
label2 &23.78 &0.69 &13.2 &21.11 &\textbf{0.73} &13.44 &\textbf{15.47} & 0.71 &\textbf{14.23} &43.04 &0.61 &11.16 &30.91 & \textbf{0.64} &\textbf{12.44} & \textbf{29.42} & 0.62 & 11.68 \\
label3 &19.95  &0.72 & 13.77 & 24.86 & 0.69 &13.03 &\textbf{12.69} & \textbf{0.75} &\textbf{15.07} &40.47 &0.61 &11.33 &27.20 & 0.67 &12.75 &\textbf{21.26} &\textbf{0.75} & \textbf{13.48} \\
label4 &33.50 &0.53 &12.28 &36.40 &0.55 &11.74 &\textbf{11.80} &\textbf{0.74} &\textbf{15.56} &52.93 &0.43 &10.87 &48.99 &0.41 &10.91 & \textbf{20.47} &\textbf{0.73} &\textbf{13.67} \\
label5 &17.99 &0.69 &14.16 &23.17 &0.56 &11.31 &\textbf{12.70}&\textbf{0.71} &\textbf{15.11} &28.73 &\textbf{0.65} &\textbf{12.51} &30.91 &0.55 &12.26 &\textbf{21.92} &0.37 &9.77 \\
label6 & 30.59 &0.60 &12.82 &35.48 &0.63 & 12.21 &\textbf{13.52} &\textbf{0.74} &\textbf{15.18} &44.54 &0.60 &11.52 &43.13 &0.52 &11.59 &\textbf{21.10}& \textbf{0.75} &\textbf{13.77} \\
label7 &23.44 &0.62 &13.64 &17.76 &0.72 &14.22 &\textbf{10.22} &\textbf{0.76} &\textbf{16.03} &28.41 &0.7 &13.02 &23.70 &0.64 &13.41 &\textbf{18.06} &\textbf{0.74} &\textbf{14.22} \\
\hline
\end{tabular}
    \begin{tablenotes}   
        \footnotesize 
        \item[1] A lower FID is better, while higher SSIM and PSNR  are more desirable.
        \item[2] Bold text indicates the best-performing values within a set of experiments.
      \end{tablenotes} 
\end{threeparttable}
\label{Font_Style_Transfer_table}
\end{table}

\begin{figure}[H]
    \centering
   \includegraphics[width=0.58\textwidth, trim=0cm 0cm 0cm 0cm, clip]{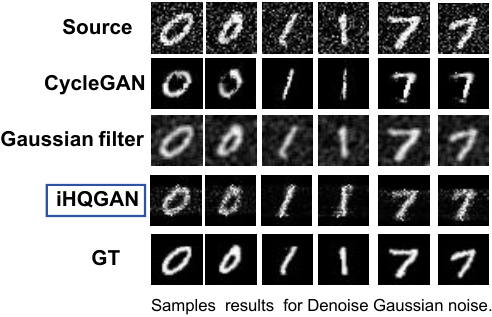}
    \caption{
     Qualitative comparison 
    of outputs generated by various methods
    trained on \textit{Image Denoising} datasets. From top to bottom:
    Source, CycleGAN, Gaussian filter, iHQGAN, and GT (ground truth).iHQGAN achieves denoising while maintaining structural consistency between the generated and source images. CycleGAN also performs denoising but fails to preserve the structural consistency between the generated and source images. The Gaussian filter's denoising effect is poor, leading to blurry images.
    }
    \label{Image_Denoising}
\end{figure}

\textbf{Image Denoising} 
As shown in Fig.\ref{Image_Denoising}, iHQGAN further demonstrates its strong generalization by effectively addressing image denoising. It outperforms classical methods such as CycleGAN and the Gaussian filter. Although CycleGAN can denoise and produce clean images, it often fails to maintain the structural consistency that is crucial for image quality.
On the other hand, the Gaussian filter can cause image blurring, compromising the clarity and sharpness of the denoised results.
Quantitative evaluations are reported in Table.~\ref{Image_Denoising_table}.
The optimal values are all concentrated in iHQGAN, demonstrating its superiority over the other two methods.

\begin{table}[h]
\belowrulesep=0pt
\aboverulesep=0pt
\centering
\caption{
Quantitative evaluation of image quality with baselines for \textit{Image Denoising}.
}
\fontsize{7}{12}\selectfont
\renewcommand{\arraystretch}{0.8}
\setlength{\tabcolsep}{0.1pt}
\begin{threeparttable}     
\begin{tabular*}{0.94\linewidth}{|@{\extracolsep{\fill}}c|ccc|ccc|ccc|@{}}
\hline
\multirow{2}{*}{Dataset}& \multicolumn{3}{c|}{CycleGAN} &\multicolumn{3}{c|}{One2One} &\multicolumn{3}{c|}{iHQGAN}  \\

\cmidrule(lr){2-4}\cmidrule(lr){5-7} \cmidrule(lr){8-10} 

&\tnote{1} FID&\tnote{1} SSIM&\tnote{1} PSNR &FID & SSIM& PSNR &FID &SSIM & PSNR \\
\hline

lable0 &16.94 &0.72 &15.44
       &21.63 &0.62 &14.44 
       &\tnote{2}\textbf{11.04} &\textbf{0.76} &\textbf{15.91}   \\     
lable1 & 9.35 & 0.63 & 18.21
       & 15.75 & 0.29 & 14.16
       &\textbf{6.37} &\textbf{0.71}&\textbf{18.94} \\
       
lable7 &18.03 &0.57 &15.53
       &17.12 &0.55 &15.75
       &\textbf{9.50} &\textbf{0.72} &\textbf{16.79 }\\
\hline
\end{tabular*}
    \begin{tablenotes} 
        \footnotesize 
        \item[1] A lower FID is better, while higher SSIM and PSNR  are more desirable.
        \item[2] Bold text indicates the best-performing values within a set of experiments.    
      \end{tablenotes} 
\end{threeparttable}
\label{Image_Denoising_table}
\end{table}
The details of the training are displayed in \ref{appendixD}.
Additionally, we observed that the images generated by iHQGAN in Fig.\ref{Edge_Detection}, Fig.\ref{Font_Style_Transfer}, and  Fig.\ref{Image_Denoising} exhibit lower brightness, which indicates that iHQGAN has a limited ability to learn brightness distributions. This issue arises because the quantum circuit structure of iHQGAN is based on that of PQWGAN \cite{tsang2023hybrid}. As previously noted in PQWGAN, an insufficient number of patches can result in darker output images. Although we used as many patches as possible, the issue remains unresolved.

\subsubsection{Analysis of the unidirectional cycle consistency constraint}
\label{Analysis of the unidirectional cycle consistency constraint}
\begin{figure}[H]
    \centering
    \begin{subfigure}{\textwidth}
        \centering
    \begin{subfigure}{\textwidth}
        \centering      
        \includegraphics[width=0.8\textwidth, trim=0cm 0cm 0cm 0cm, clip]{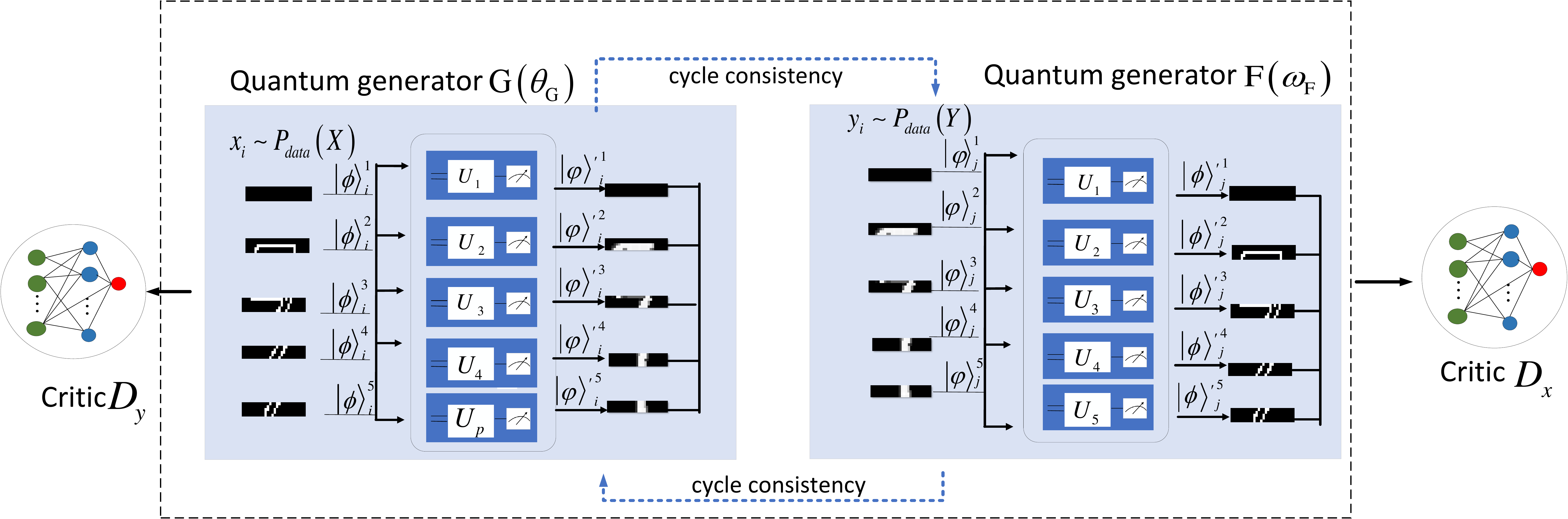}
        \caption*{\footnotesize (a)Q-wcycleGAN.
       It features two quantum generators with the same structure and unshared parameters,
       implementing bidirectional cycle consistency constraints between the two quantum generators. 
        }
        \label{subfig:subfig1}
    \end{subfigure}
    \caption{Two different quantum schemes utilize bidirectional cycle consistency constraints.} 
    
    \includegraphics[width=0.8\textwidth, trim=0cm 0cm 0cm 0cm, clip]{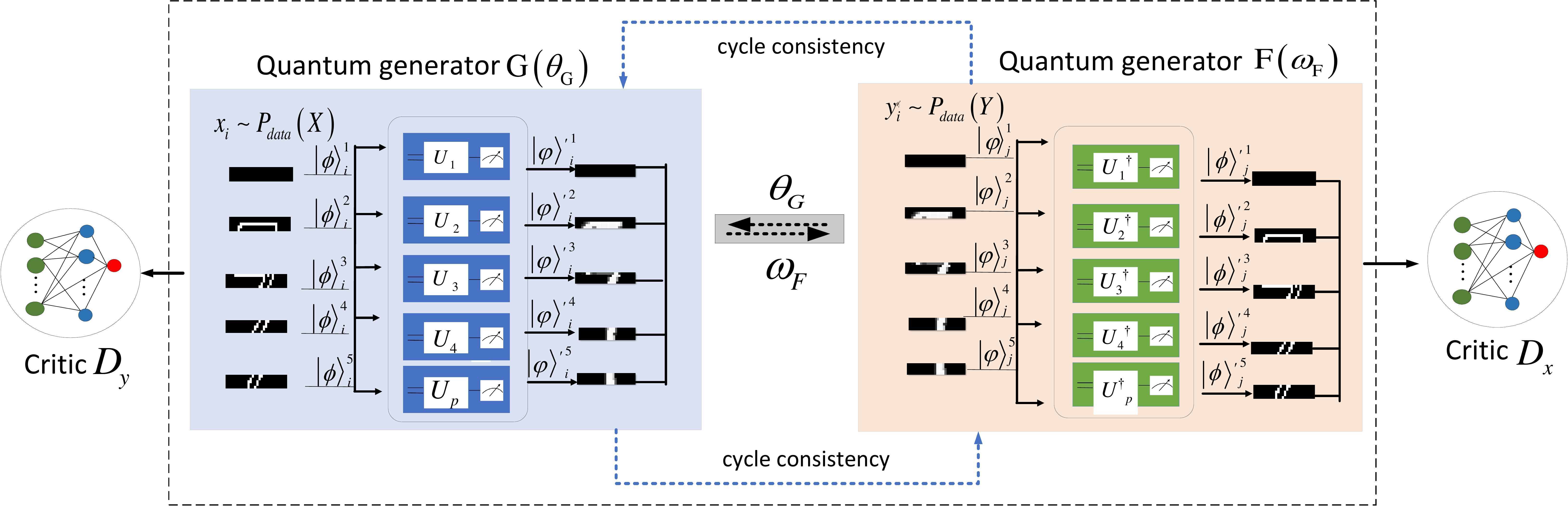}
        \caption*{\footnotesize (b)
        iHQGAN w/o ACNNs. It includes two parameter-sharing and mutually reversible quantum generators, implementing bidirectional cycle consistency constraints between the quantum generators
       
        }
        \label{subfig:subfig2}
    \end{subfigure}
    \label{Architecture}
\end{figure}

iHQGAN employs a unidirectional cycle consistency constraint between each quantum generator and its
corresponding ACNN.
We compared iHQGAN with two quantum schemes that utilize bidirectional cycle consistency constraints between two quantum generators: Q-wcycleGAN and iHQGAN w/o ACNNs. The two quantum schemes are illustrated in Fig.\ref{Architecture}. Experiments were conducted on the sub-dataset  with label 0 from the \textit{Edge Detection} dataset.

\begin{figure}[H]
    \centering
   \includegraphics[width=0.6\textwidth, trim=0cm 0cm 0cm 0cm, clip]{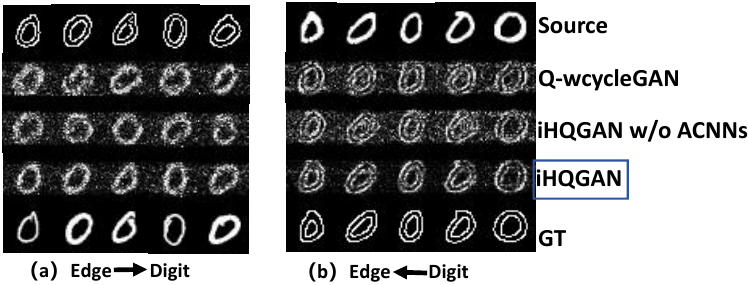}
    \caption{\footnotesize 
     Qualitative comparisons of outputs generated by various quantum schemes on the sub-dataset with label 0 from the \textit{Edge Detection} dataset.The blue solid boxes highlight the key experiment.
    } 
    \label{Architecture_result_image}
\end{figure}

The qualitative and quantitative comparisons of the above two schemes with iHQGAN are displayed in Fig.\ref{Architecture_result_image} and Table.~\ref{tab_architecture}.
iHQGAN produces clearer, higher-fidelity images in both directions. Compared with iHQGAN, Q-wcycleGAN and iHQGAN w/o ACNNs produce images with more noise.
Q-wcycleGAN struggles to maintain structural consistency in both directions, while iHQGAN w/o ACNNs maintains it only in the Edge $\longleftarrow$ Digit direction.
Table.~\ref{tab_architecture} shows that iHQGAN surpasses the other methods across all metrics for both directions, highlighting its superior image data modeling, structural consistency preservation, and image quality enhancement.
The superior performance of iHQGAN can be attributed to the introduction of ACNNs through unidirectional cycle consistency constraints, which help the quantum generators find the optimal parameter space during training.
In contrast, Q-wcycleGAN and iHQGAN w/o ACNNs encounter challenges in optimizing bidirectional cycle consistency due to the limited expressive capability of quantum generators. 
These results indicate that the bidirectional cycle consistency strategy fails to maintain structural consistency in both directions simultaneously.

\begin{table}[h]
\caption{
Quantitative comparison of outputs generated by various quantum schemes.
}
\fontsize{7}{12}\selectfont
\label{tab_architecture}
\renewcommand{\arraystretch}{0.8} 
\setlength{\tabcolsep}{3pt} 
\begin{threeparttable} 
\begin{tabular*}{1.0\linewidth}{@{\extracolsep{\fill}}cccccccc@{}}
\hline
\multirow{2}{*}{Method} & \multicolumn{3}{c}{ Edge$\longrightarrow$Digit} & \multicolumn{3}{c}{ Edge$\longleftarrow$Digit} \\
\cmidrule(lr){2-4} \cmidrule(lr){5-7}  
 & \tnote{1} FID &  \tnote{1} SSIM& \tnote{1} PSNR
 & FID & SSIM& PSNR
 \\ 
\hline
Q-wcycleGAN&23.94&0.39
&10.5&29.13&0.3&9.9
 \\
iHQGAN w/o ACNNs &26.30&0.35
&10.19 &31.17 &0.32 &10.0
\\
iHQGAN &\tnote{2} \textbf{22.46}&\textbf{0.59}
&\textbf{12.53} &\textbf{18.95} &\textbf{0.5}&\textbf{11.64}
 \\
\hline
\end{tabular*}
\begin{tablenotes} 
        \footnotesize 
      \item[1] A lower FID is better, while higher SSIM and PSNR are more desirable.
        \item[2] Bold text indicates the best-performing parameters within a set of experiments.
        
      \end{tablenotes} 
\end{threeparttable}
\end{table}

\subsubsection{An analysis of the hyperparameters in the loss function of the quantum generators}
\label{Analysis of hyperparameters}
The loss function of the quantum generators ,as seen in Equation \ref{Loss:G} and Equation \ref{Loss:F} ,
draws on that of  classical generators. 
In GANs, the learning rates of the classical generator and classical discriminator are similar, and the related work sets the hyperparameters in the loss function of the classical generators  to  $\varepsilon = 1,\eta = 10, \rho= 150$, where $\rho = 15\eta $ \cite{chen2019quality}.
However, QGANs often feature a higher learning rate for the quantum generator compared to the classical discriminator.\cite{huang2021experimental,vieloszynski2024latentqganhybridqganclassical}. 
Therefore, we need to find an appropriate combination of hyperparameters.
Experiments were conducted on 
the sub-dataset  with label 0 from the \textit{Edge Detection} dataset 
We set the default values for the parameter pair as $ (\eta, \rho) $ are $\eta = 10$ and $\rho = 150$.
Specifically, we explored $\varepsilon = 10 $ in combination with various multiples of the parameter pair $ (\eta, \rho) $, and  $\varepsilon = 20$  with  $ 3 \times (\eta, \rho)$. 
\begin{figure}[H]
    \centering
   \includegraphics[width=0.5\textwidth, trim=0cm 0cm 0cm 0cm, clip]{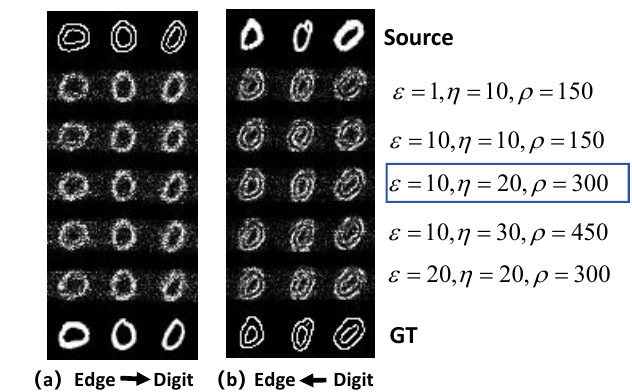}
    \caption{\footnotesize 
     Qualitative comparisons of outputs generated by different hyperparameter combinations of iHQGAN on the sub-dataset  with label 0 from the \textit{Edge Detection} dataset.
     The related work in classical fields sets $\varepsilon= 1, \eta = 10, \rho = 150$, and $\rho = 15\eta$. We set the default values for the parameter pair $ (\eta, \rho) $ as $\eta = 10$ and $\rho = 150$. For iHQGAN, we explored $\varepsilon= 10 $ with various multiples of the parameter pair $ (\eta, \rho) $,as well as $\varepsilon = 20 $ with $3 \times (\eta, \rho) $. The blue solid boxes highlight the key experiment.
    } 
    \label{Hyperparameter1}
\end{figure}

\begin{figure}[H]
    \centering
   \includegraphics[width=1\textwidth, trim=0cm 0cm 0cm 0cm, clip]{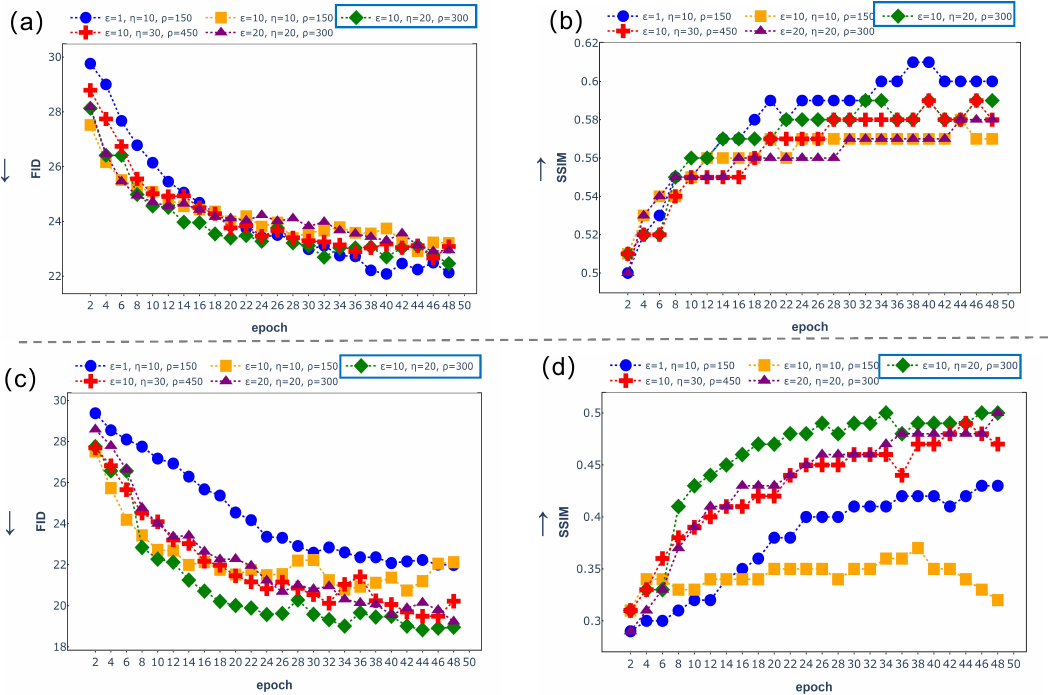}
    \caption{\footnotesize 
    Quantitative comparison of iHQGAN's performance across different hyperparameter combinations on the test set, evaluated over 50 training sessions on the sub-dataset  with label 0 from the \textit{Edge Detection} dataset.
    Subplots (a) and (b) above the dashed line present the performance in the Edge $\longrightarrow$ Digit direction. Subplots (c) and (d) below the dashed line illustrate the performance in the Edge $\longleftarrow$ Digit direction.The blue solid boxes highlight the key experiment. A lower FID is better, while a higher SSIM is more desirable.
    }
    \label{Hyperparameter2}
\end{figure}

Fig.\ref{Hyperparameter1} presents qualitative experimental results, 
while Fig.\ref{Hyperparameter2} quantitatively compares iHQGAN's performance across various hyperparameter combinations on the test set, evaluated over 50 training sessions using the sub-dataset  with label 0 from the \textit{Edge Detection} dataset. The images generated with the combination $\varepsilon = 10, \eta = 20$, and $\rho= 300$  show clearer and more distinct contours,  yielding stable and excellent FID and SSIM scores in both directions. In contrast, the combination  $\varepsilon = 1, \eta = 10$, and $\rho= 150$ produces somewhat blurry contours in the generated images in the Edge $\longleftarrow$ Digit direction, as evidenced by the corresponding poor FID and SSIM scores. 
Additionally, images produced with the combination  $\varepsilon = 10, \eta = 10$, and $\rho= 150$ fail to  maintain structural consistency in the Edge $\longleftarrow$ 
Digit direction, with the SSIM scores reflecting inadequate performance.
The remaining three combinations produce visually similar images, and their values are close around the 50th epoch.
Our results indicate that when the weight  $\varepsilon$  of the adversarial losses is too low relative to the weight $\rho$ of the quality-aware loss, as in the combination  $\varepsilon = 1, \eta = 10$, and $\rho= 150$, the learning of the image data distribution weakens, resulting in blurred contours. Conversely, when the weight $\varepsilon$ of the adversarial losses is relatively high compared with that of the quality-aware loss,
as seen in the combination $\varepsilon = 10, \eta = 10$, and $\rho= 150$, it compromises the structural consistency between the generated images and the source images in
a certain translation direction.  

\section{Discussion and Further Work}
\label{Discussion and Further Work}
This paper combines the invertibility of quantum computing with the approximate reversible nature of unsupervised I2I translation and proposes iHQGAN, a lightweight unsupervised I2I translation framework. 
iHQGAN is the first versatile quantum solution for unsupervised I2I translation, which
extends QGAN research into more complex image generation scenarios and offers a quantum approach that reduces the parameters of GAN-based unsupervised I2I translation methods.
Compared with classical GAN-based methods for unsupervised I2I translation \cite{isola2017image,chen2019quality,YunjeyStarGAN}, this study introduces quantum technology to perform this task. 
Furthermore, similar to classical reversible methods (\cite{van2019reversible,dai2021iflowgan}, which reduce parameters using a classical reversible mechanism, iHQGAN achieves parameter reduction through a quantum reversible mechanism.
Currently, QGAN is limited to generating single-channel,low-resolution images, such as MNIST and MNIST-C. To assess the model's performance, this study processed the MNIST dataset to obtain three unsupervised I2I translation tasks,with a total of 19 sub-datasets.

In Section \ref{Experiment on different unsupervised I2I translation tasks}, 
the experiments conducted on the \textit{Edge Detection} and \textit{Font Style Transfer} datasets
demonstrated that iHQGAN can effectively 
achieve unsupervised I2I translation and
reduce the parameter scale.
As can be seen from the qualitative evaluation in Fig.\ref{Edge_Detection} and Fig.\ref{Font_Style_Transfer}, the generated images of iHQGAN achieve domain transfer while maintaining structural consistency with the source images. In contrast, the classical irreversible method CycleGAN and the classical reversible method One2One using low-complexity CNN-based generators do not perform better than iHQGAN.Many generated images of the latter classical methods exhibit pixel loss and 
fail to maintain structural consistency with the source images.
Table.~\ref{Edge_Detection_table} and Table.~\ref{Font_Style_Transfer_table} quantitatively show that many best values are concentrated in iHQGAN. 
Additionally, as shown in Table.~\ref{Parameter_counts_table}, 
like classical reversible methods, iHQGAN leverages a reversible mechanism to reduce the parameters of classical irreversible methods.
In fact, many classical reversible methods leverage the 
approximate reversible nature of unsupervised I2I translation to construct reversible models for reducing parameter counts.\cite{shen2020one,dai2021iflowgan,van2019reversible} 
Furthermore, we conducted experiments on \textit{Image Denoising} to validate the model's generalization ability. 
As shown in  Fig.\ref{Image_Denoising} and Table.~\ref{Image_Denoising_table},the results indicate that, compared with the classical nonlinear,low-complexity CNN-based CycleGAN and the classical linear Gaussian filter, the denoised images produced by iHQGAN demonstrate significant denoising effectiveness and higher fidelity.
The effectiveness of the iHQGAN framework can be attributed to the interplay of various mechanisms, including the unidirectional cycle consistency constraint, the appropriate loss function, two mutually reversible generators with shared parameters, and the alternating training strategy.

In Section \ref{Analysis of the unidirectional cycle consistency constraint}, 
our work
compared the effectiveness of the strategy of unidirectional  cycle consistency constraints  between each quantum generator and its corresponding ACNN in iHQGAN with
bidirectional cycle consistency constraints between two quantum generators. We designed two other quantum schemes adopting bidirectional cycle consistency constraints for unsupervised I2I translation, as illustrated in Fig.\ref{Architecture}.
All quantum schemes
were tested on the sub-dataset with label 0 from the \textit{Edge Detection} dataset.
Fig.\ref{Architecture_result_image} indicates that, compared with iHQGAN,  Q-wcycleGAN and iHQGAN w/o ACNNs fail to maintain structural consistency simultaneously in both directions. Neither generator in Q-wcycleGAN achieves this consistency, while iHQGAN w/o ANNs has only one generator that does. As can be seen from  Table.~\ref{tab_architecture},
all best values are concentrated in iHQGAN.
This is because unidirectional cycle consistency constraints enable ANNs to optimize the parameter space of the quantum generators.
In the classical domain, the strategy of utilizing unidirectional cycle consistency constraints can achieve unsupervised image translation \cite{fu2019geometry,benaim2017one}, and numerous studies have demonstrated that classical components can enhance the performance of quantum circuits \cite{callison2022hybrid,chang2024latent,shu2024variational}.

In Section \ref{Analysis of hyperparameters}, we investigated the effect of different hyperparameters combinations in the loss function of the quantum  generators.The loss function of the quantum  generators ,as defined in Equation \ref{Loss:G} and Equation \ref{Loss:F},draws on that of classical generators. 
In GANs, the learning rates of the classical generator and classical discriminator are close.\cite{chen2019quality}.
However, the learning rate of the classical critic is relatively lower than that of the quantum generator in QGANs \cite{huang2021experimental,tsang2023hybrid,vieloszynski2024latentqganhybridqganclassical} .Therefore,it is crucial to explore appropriate hyperparameters combinations of the loss function for iHQGAN.
Experiments were conducted  on the sub-dataset with label 0 from the \textit{Edge Detection} dataset.
Fig.\ref{Hyperparameter1} and Fig.\ref{Hyperparameter2} 
present qualitative and quantitative comparisons of the results under different weight combinations.
When the weight of the adversarial losses is too low relative to that of the IQA loss in the loss function, the learning of the image data distribution weakens, resulting in blurred contours.  Conversely, 
it compromises the structural consistency between  the generated images and the source images in a certain translation direction.

This groundbreaking work paved the way for quantum technology to address complex image translation tasks.
However, the quantum generator's structure still requires improvement, as image quality depends on its expressive capacity.
This study draws on the quantum generator's structure from PQWGAN \cite{tsang2023hybrid}. However, the images generated by PQWGAN suffer from many flaws, such as poorly fitting luminance distributions and excessive noise.
We plan to explore how to design quantum generators with better expressiveness to improve the quality of generated images from the perspective of the number of qubits, the depth of quantum circuits, and efficient ansatz designs, which stands as a significant future endeavor within our research trajectory.
\section{Conclusion}
\label{Conclusion}
This paper presents iHQGAN, the first versatile quantum method for unsupervised I2I translation. iHQGAN combines the approximate reversibility of unsupervised I2I translation with the invertibility of quantum computing, extending the application of QGANs to complex image generation scenarios and effectively addressing the issue of excessive parameters in classical GAN-based unsupervised I2I translation methods. Simulation experiments on extensive datasets demonstrate that iHQGAN effectively performs unsupervised I2I translation with strong generalization ,and can outperform  classical methods that use low-complexity CNN-based generators.
Additionally, iHQGAN, like classical reversible methods, reduces
the number of parameters in  classical irreversible methods via a reversible mechanism.
These results highlight the effectiveness, broad applicability, and parameter-saving nature
of the iHQGAN in unsupervised I2I translation.
Future research will revolve around developing quantum generators with powerful expressiveness, aiming to improve the overall quality of images generated by iHQGAN.

\section*{Authors' contributions}
\textbf{Xue Yang}: Conceptualization, Methodology, Validation, and writing the article. \textbf{Ri-Gui Zhou}: Project administration, Supervision. \textbf{Yao-Chong Li}: Conceptualization. \textbf{Shi-Zheng Jia}: Conceptualization, Software. \textbf{Zheng-Yu Long}: Reviewing. \textbf{Cheng-Ji Yan} Reviewing.\textbf{Wen-Shan Xu}:Reviewing. \textbf{Wen-Yu Guo}: Formal analysis. \textbf{Fu-Hui Xiong}: Formal analysis.

\section*{Availability of supporting data}
All data generated or analysed during this study are available and included in this published article. \href{https://github.com/yxSMU/iHQGAN}{Code} for our work has been open-sourced.

\section*{Acknowledgements}
This work is supported by the National Natural Science Foundation of China under Grant No. 62172268 and 62302289; Shanghai Science and Technology Project under Grant No. 21JC1402800 and 23YF1416200.
\small
\bibliographystyle{elsarticle-harv}
\bibliography{iHQGAN}

\newpage

\appendix 

\section{Details of quantum circuits of iHQGAN}
\label{appendixA}

\begin{figure}[H]
	\centering
	\includegraphics[width=0.68\textwidth]{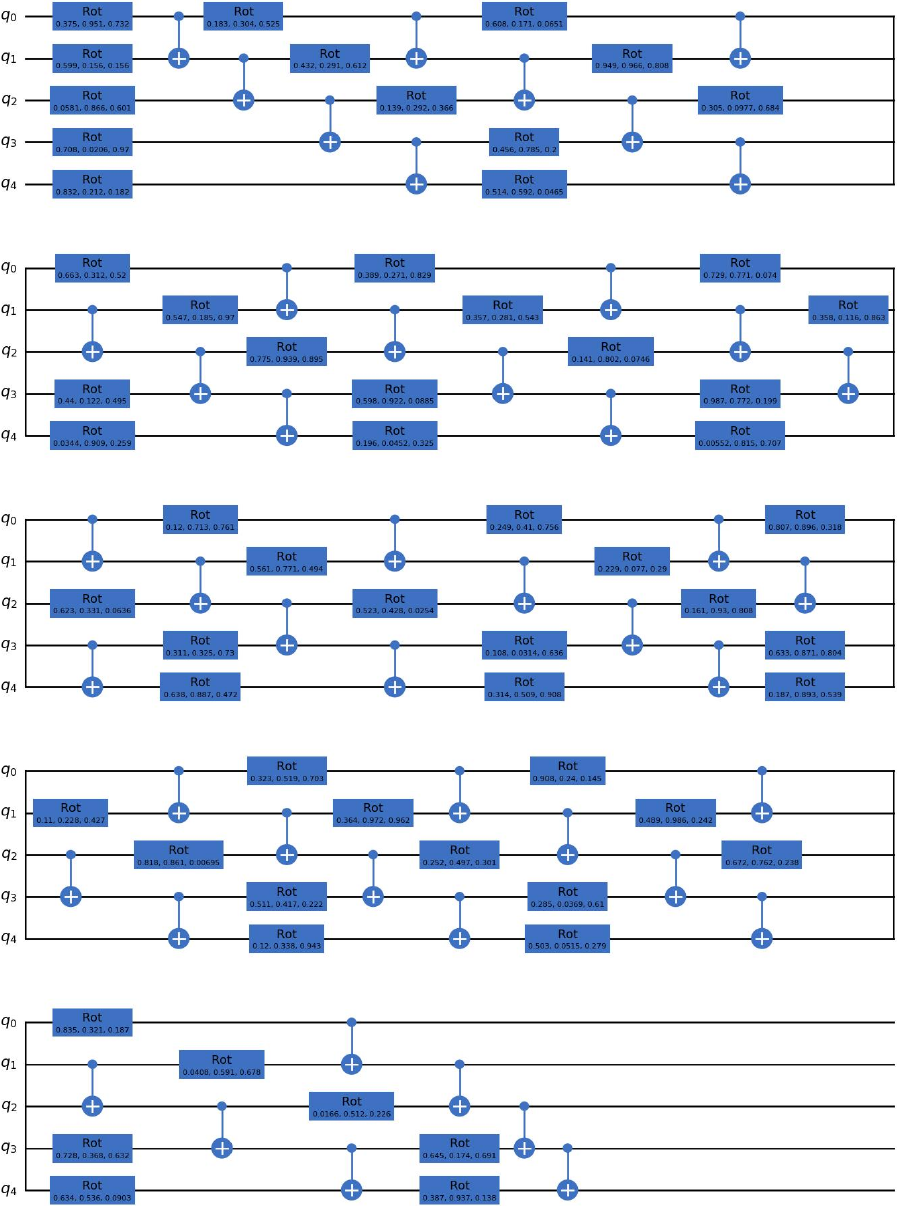}
	\caption{
		\footnotesize
		The quantum circuits $U_{k} (k=1...p)$ of the quantum generator $G$ feature a depth of 24 and consist of 48 CNOT operations.}
\end{figure}

\begin{figure}[H]
	\centering
	\includegraphics[width=0.78\textwidth]{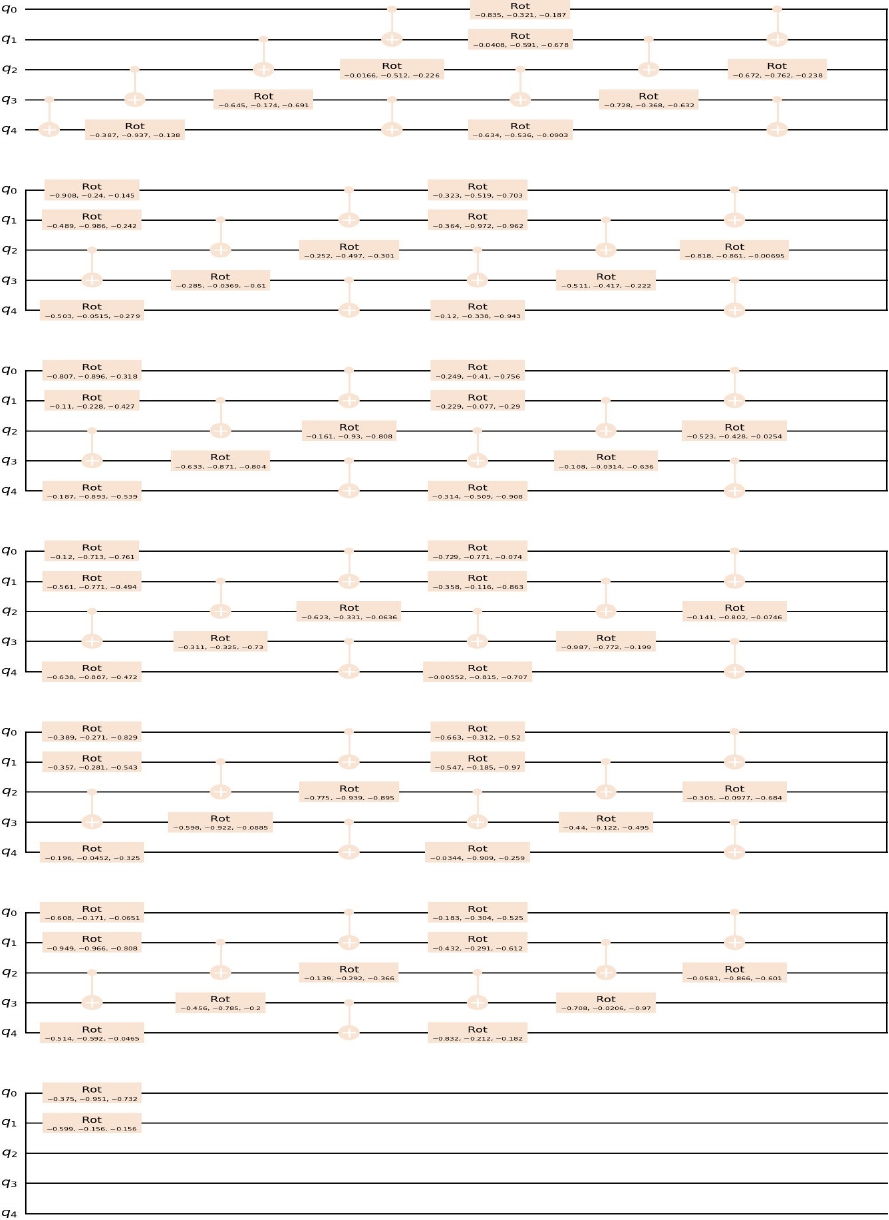}
	\caption{
		\footnotesize
        The quantum circuits $U_{l} (l=1...p)$ of the quantum generator $F$ feature a depth of 24 and consist of 48 CNOT operations.
  }
\end{figure}

\section{Architecture of the classical critic }
\label{appendixB}
The critics  $D_X$ and $D_Y$  are classical neural networks, with parameters $\kappa_{D_{\mathrm{x}}}$ and $\kappa_{D_{\mathrm{Y}}}$ , respectively. 
Their structure is the same as that of the critic in PQWGAN, designed to assist in training the quantum generator. After training is complete, the critic  is discarded. One of the important principles of training an effective QGAN model is to ensure that the generator and the critic  have comparable capabilities.
Mode collapse may occur if the generator is overpowering and the critic  is weak. Conversely, it is difficult for the generator to produce realistic samples.
There is no way to accurately determine the relationship between the learning ability of the classical critic  and the quantum generator. This paper draws on the PQWGAN experience where the critic  is a fully connected network with two hidden layers of 1024, 512, 256, and 1 neuron. 
Both hidden layers use a leakyReLU activation function with a slope of 0.2, which introduces more nonlinearities to increase the expressive power of the neural network. 
The final hidden layer connects to the output layer, which has only one neuron to produce a real value and determine whether the input data is real or synthetic.

\section{How to build datasets}
\label{appendixC}
The MNIST-C dataset is a benchmark for robustness in computer vision. \textit{Image Denoising} and \textit{Edge Detection} correspond to the two corruptions - Gaussian Noise, and Canny Edges in  MNIST-C code. For \textit{Font Style Transfer tasks}, we applied dilation to the MNIST dataset using PyTorch and OpenCV libraries.
First, the MNIST dataset is loaded using PyTorch's \texttt{datasets.MNIST } function, followed by converting the images into NumPy arrays.
Subsequently, a 2x2 dilation kernel is defined, and the dilation operation is executed using OpenCV's \texttt{cv2.dilate} function. Finally, the data are converted back to PyTorch tensors and stored in PNG format.
We have published the dataset-making code in the GitHub repository.

\section{The display of more training details}
\label{appendixD}
\begin{figure}[H]
	\centering
	\includegraphics[width=0.7\textwidth]{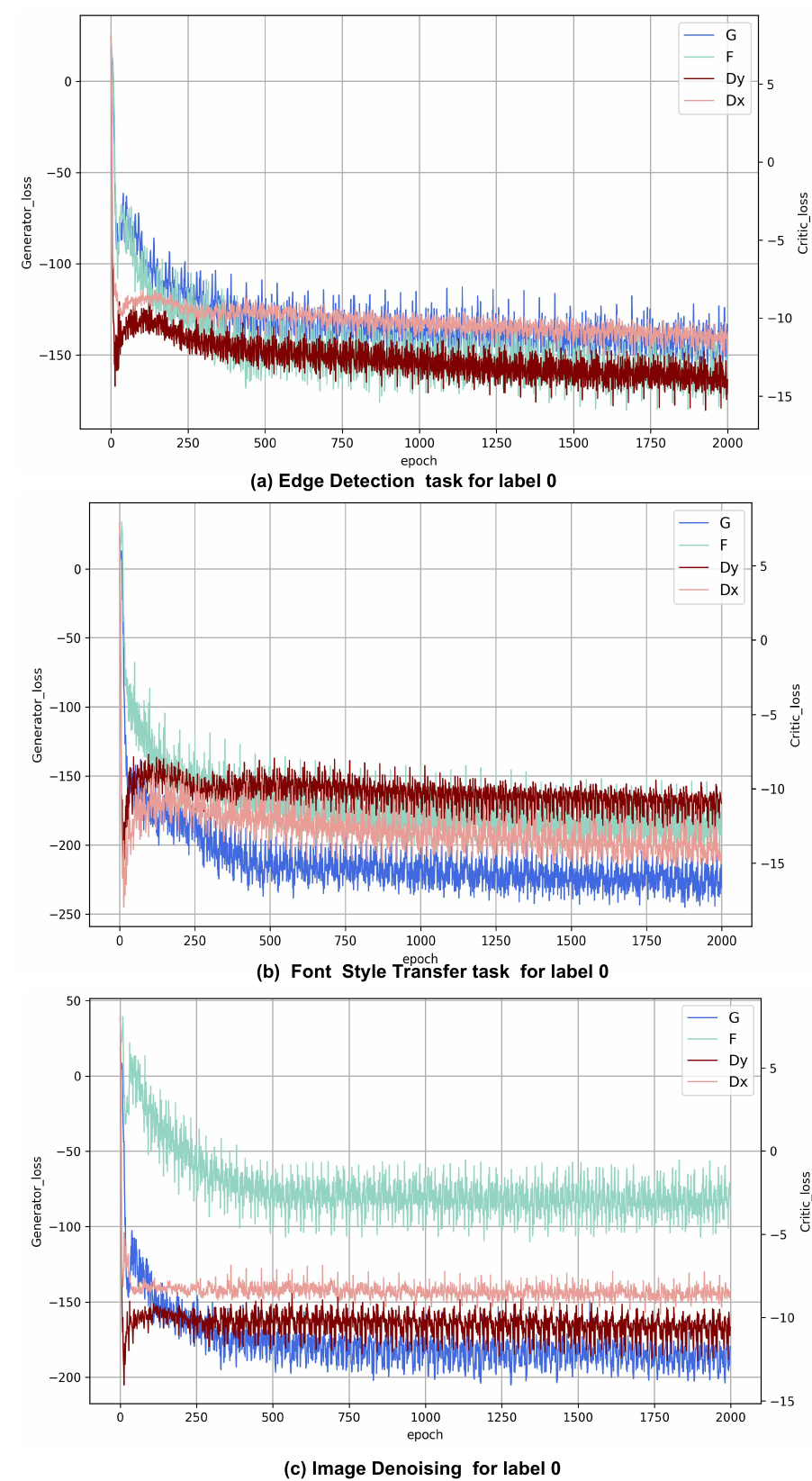}
	\caption{
		\footnotesize
               Training loss curves on three sub-datasets with label 0 from different training datasets.
		}
         \label{DF1}
\end{figure}

\begin{figure}[H]
	\centering
	\rotatebox{90}{\includegraphics[width=1.1\textwidth]{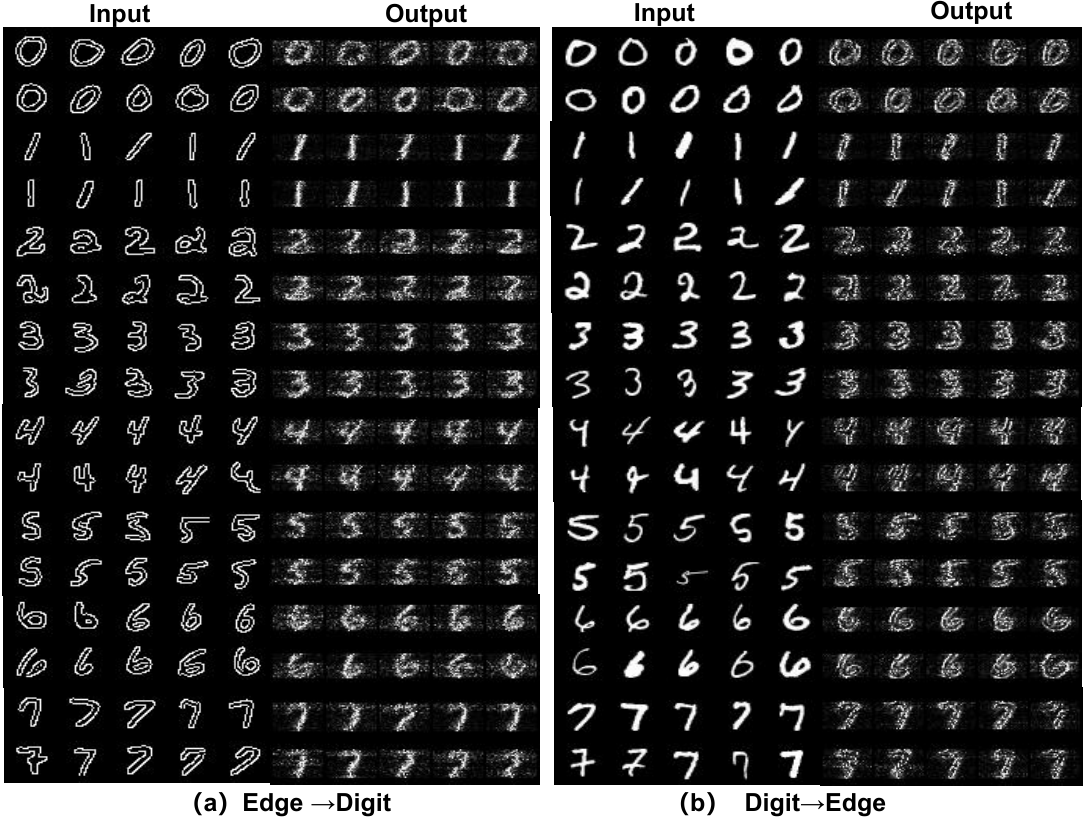}}
	\caption{
		\footnotesize
           The examples of training results of \textit{Edge Detection}. Sub-figure(a) shows the examples in the Edge $\rightarrow$ Digit direction. Sub-figure(b)  the examples in the Digit $\rightarrow$ Edge direction
  }
  \label{DF2}
\end{figure}

\begin{figure}[H]
	\centering
	\rotatebox{90}{\includegraphics[width=1.1\textwidth]{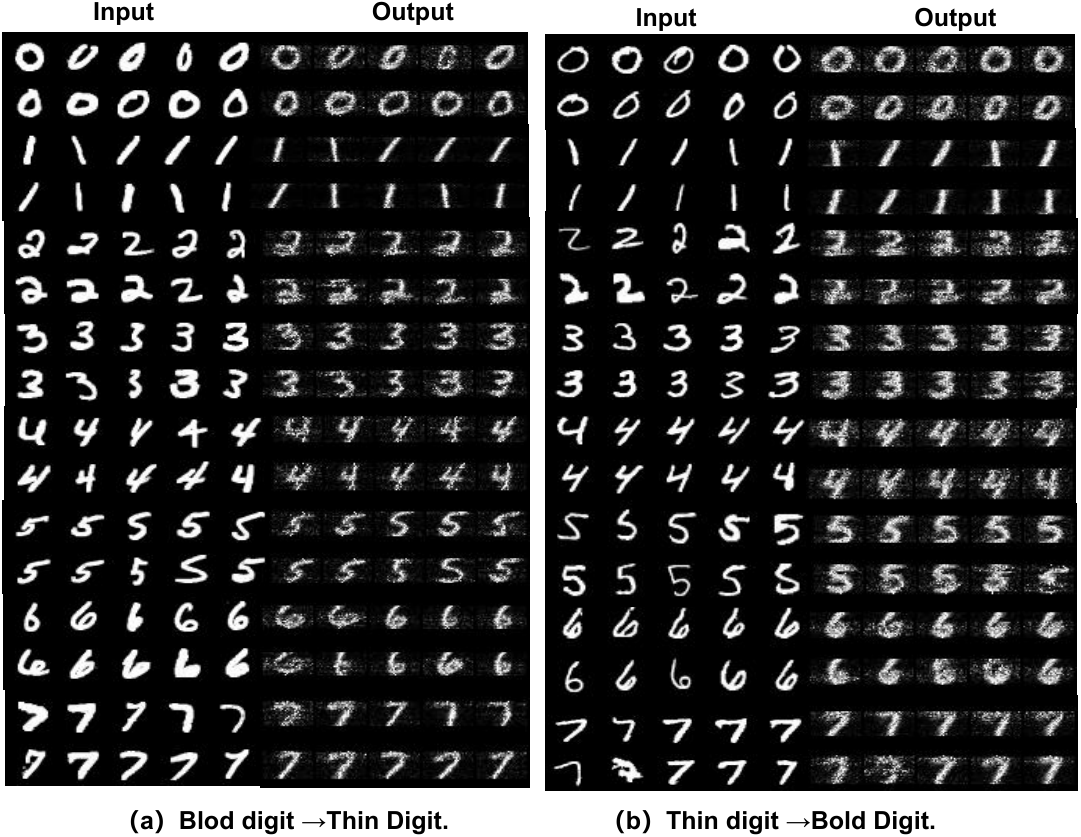}}
	\caption{
		\footnotesize
		The examples of training results of \textit{Font Style Transfer}. Sub-figure(a) shows the examples in the Bold Digit $\rightarrow$  Thin Digit direction. Sub-figure(b) shows  the examples in the Thin Digit $\rightarrow$ Bold Digit direction}
       \label{DF3}
\end{figure}

\begin{figure}[H]
    \centering
    \includegraphics[width=0.8\textwidth, trim=0 12 0 0cm, clip]{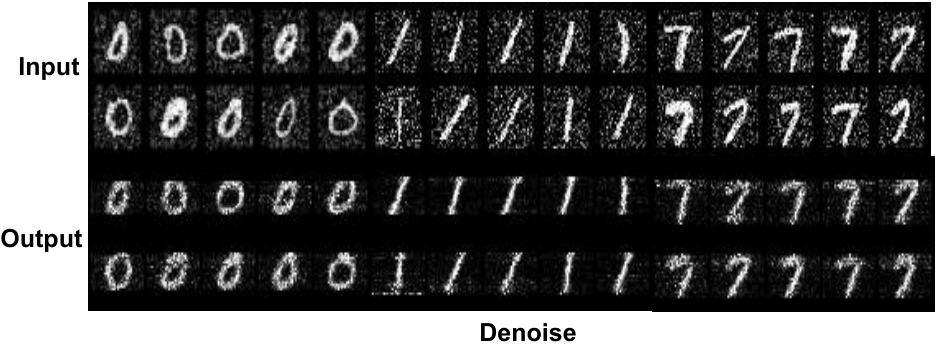}
    \caption{\footnotesize The examples of training results of \textit{Image Denoising}}
    \label{DF4}
\end{figure}

Fig.\ref{DF1} shows the loss curves on three sub-datasets with label 0 from different training datasets.
The generator loss initially decreases and then stabilizes during training, whereas the critic  loss first increases before stabilizing, ultimately reaching a Nash equilibrium.
Fig.\ref{DF2},Fig.\ref{DF3} and Fig.\ref{DF4},show examples of training results for \textit{Edge Detection}, and \textit{Font Style Transfer },\textit{Image Denoising}. 
iHQGAN can effectively perform unsupervised I2I translation across the three tasks, demonstrating stable performance and impressive generalization.

\end{document}